# Scintillations in Southern Europe during the geomagnetic storm of June 2015: analysis of a plasma bubbles spill-over using ground-based data

Anna Morozova [1,2,*], Luca Spogli [3], Teresa Barata [1,4], Rayan Imam [3], Emanuele Pica [3], Juan Andrés Cahuasquí [5], Mohammed Mainul Hoque [5], Norbert Jakowski [5] and Daniela Estaço [6]

[1] Instituto de Astrofísica e Ciências do Espaço, University of Coimbra, 3004-531 Coimbra, Portugal; anna.morozova@uc.pt (A.M.), teresabarata@dct.uc.pt (TB)
[2] Department of Physics, FCTUC, University of Coimbra, 3004-531 Coimbra, Portugal
[3] Istituto Nazionale di Geofisica e Vulcanologia, Rome, Italy; luca.spogli@ingv.it (LS), rayan.imam@ingv.it (RI), emanuele.pica@ingv.it (EP)
[4] Department of Earth Sciences, FCTUC, University of Coimbra, 3004-531 Coimbra, Portugal
[5] Institute for Solar-Terrestrial Physics, German Aerospace Center DLR, Neustrelitz, Germany; Andres.Cahuasqui@dlr.de (AC), Mainul.Hoque@dlr.de (MH), Norbert.Jakowski@dlr.de (NJ)
[6] Departamento de Física da University of Aveiro, Aveiro, Portugal (at the time the research was conducted)
* Correspondence: anna.morozova@uc.pt or annamorozovauc@gmail.com

**Abstract:** The sensitivity of Global Navigation Satellite Systems (GNSS) receivers to ionospheric disturbances and their constant growth are nowadays resulting in an increased concern of GNSS-users about the impacts of ionospheric disturbances at mid-latitudes. The geomagnetic storm of June 2015 is an example of a rare phenomenon of a spill-over of equatorial plasma bubbles well North from their habitual. We study the occurrence of small- and medium-scale irregularities in the North Atlantic Eastern-Mediterranean mid- and low-latitudinal zone by analysing the behaviour of the amplitude scintillation index S4 and of the rate of total electron content index (ROTI) during such a storm. In addition, large scale perturbations of the ionospheric electron density were studied using ground and space-born instruments, thus characterizing a complex perturbation behaviour over the region mentioned above. The multi-source data allows us to characterize the impact of irregularities of different scales to better understand the ionospheric dynamics and stress the importance of a proper monitoring of the ionosphere in the studied region.

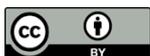

## 1. Introduction

The analysis presented here aims to investigate a rare event of a spill-over of equatorial plasma bubbles (EPBs) from low latitudes, triggered by the occurrence of a geomagnetic storm. These spill-over events are gaining more interests as they are one of the main sources of storm-trigger Global Navigation Satellite Systems (GNSS) disturbances at mid-latitudes. A first assessment of this class of disturbances on the investigated storm has been proposed by [1], who reported about large-scale EPBs events, covering the Mediterranean sector.

As highlighted by [2], the conditions favouring the spill-over of EPBs at mid-latitudes are mostly due to the occurrence of storm-induced prompt penetration electric fields - PPEF, [3] occurring near the local sunset hours at the geomagnetic equator and that may intensify the EPBs seeding process due to pre-reversal enhancement [4]. This has been also nicely illustrated during the recent May 2024 Mother's Day superstorm, during which a meaningful EPB spill-over has been recorded over the Mediterranean [5]. The authors of [6] reported a mid-latitude scintillation event over the Southern United States on June 1, 2013, occurring during the main phase of a modest magnetic storm. This suggests that the intensity of the storm is not the primary factor; rather, it is the local time of the occurrence of the PPEF event that triggers the EPB spill-over at mid-latitudes.

The geomagnetic storm of June 2015 and its effects on the ionospheric conditions are well described in previous work: see, for example, [7-9] and references therein for a detailed description of this geomagnetic storm development and ionospheric responses with a focus on the region studied in this work (the West Mediterranean, the Iberian Peninsula and a part of the Eastern North Atlantic ocean between Azores, Canary and Madeira archipelagos and the North-Western African coast, or 65° W - 65° E in longitude and 0° - 50° N in latitude, see also Fig. 1).

In brief, the storm of June 22-23, 2015, was the second strongest geomagnetic storm of the 24th solar cycle caused by three consecutive coronal mass ejections [10-11]. The response of the ionosphere during its initial phase was asymmetric concerning the hemispheres, mainly due to seasonal effects, with a negative phase of the storm observed in most of the Northern hemisphere and a positive phase in the Southern hemisphere and Western longitudes of the Northern hemisphere, over the North Atlantic Ocean - the area of our study. The total electron content (TEC) variations were obtained from Portuguese GNSS receivers located in Lisbon (Iberian Peninsula), S. Miguel Island (Azores) and the Madeira island [7, 9] and consisted of a strong positive ionospheric storm that started in the late afternoon of June 22 ($\Delta$TEC ~+20 TECu) and a strong negative ionospheric storm ($\Delta$TEC ~-20 TECu) on the following day, June 23, 2015. A secondary Dst drop on June 25, 2015, caused another positive-negative ionospheric storm observed on June 25-26 [7, 9], but the secondary event was not accompanied by scintillations and therefore is not analysed in this work.

The storm of June 22, 2015, was associated with PPEFs [12] and as was mentioned before, these conditions may be favourable both for the development of EPBs during the storm time and their spill-over to the middle latitudes. Such a rare phenomenon as EPB spill-over was observed twice during this storm: first in the Euro-African longitudinal sector between 20h UTC of June 22 and 02h UTC of June 23, and later, between 04h UTC and 07h UTC of June 23, in the American longitudinal sector [12]. The EPB spill-over resulted in strong scintillations of the GNSS signal in the middle latitudes, as, for example, is reported in [1 and 8]. In particular, the quality of the relative positioning, especially the single frequency relative positioning [8], for the area of the Iberian Peninsula decreased drastically between 22-19 h UTC of June 22 and 06-08 h UTC of June 23. Those high positioning errors were found to be coincident with periods of high spatial TEC gradient of the medium to large scales, as reported by [8] using the gradient ionosphere index (GIX).

Previously, the scintillation event of June 22-23, 2015, as related to the EPB spill-over was studied using the rate of TEC index (ROTI) [14], obtained both from the ground-based GNSS receivers and from space-born instruments [1, 8, 12-13] as well as data on spatial and temporal TEC gradients [8].

In this work we present, for the first time, an analysis of this event using the amplitude scintillation index S4 [15-16] obtained at three locations in the Mediterranean - North Atlantic area (15° N - 50° N, 25° W - 30° E). The S4 data are complemented with ROTI obtained from ground-based instruments of two kinds: geodetic receivers and ionospheric scintillation monitoring receivers (ISMRs) [17-18]. The geodetic receivers can provide Total Electron Content (TEC), and the index based on its rate of change (ROTI), while the scintillation receivers provide the amplitude scintillation index S4. Since the variations of the S4 and ROTI indices may result from ionospheric inhomogeneities of different spatial scales [19], the comparison of the S4 and ROTI data will provide an insight into the dynamics of the ionosphere during the studied event. The character of the ionospheric electron density inhomogeneities (size, location, travelling direction) was obtained from the analysis of the slant TEC (sTEC) data and the spatial TEC gradients calculated for the whole studied area using ground-based data sources and the Swarm data sets.

## 2. Data

*2.1 S4 and ROTI data*

The data on the scintillation indices variations, the amplitude scintillation index S4 and the rate of TEC index (ROTI), are obtained from a set of GNSS receivers.

2.1.1 S4 data

In the presented analysis, we focus on the S4 index computed over 1-min intervals, able to reveal the impact of small-scale irregularities on the received signal. These have a typical scale size below the Fresnel's scale for L-band signals and GNSS observational geometry from ground, i.e., a few hundred meters scale [20].

Data from three receivers were used to analyse S4 variations during the studied event. The original data have 1-min time resolution and obtained using GPS satellites only. These receivers are marked as open red diamonds in Fig. 1.

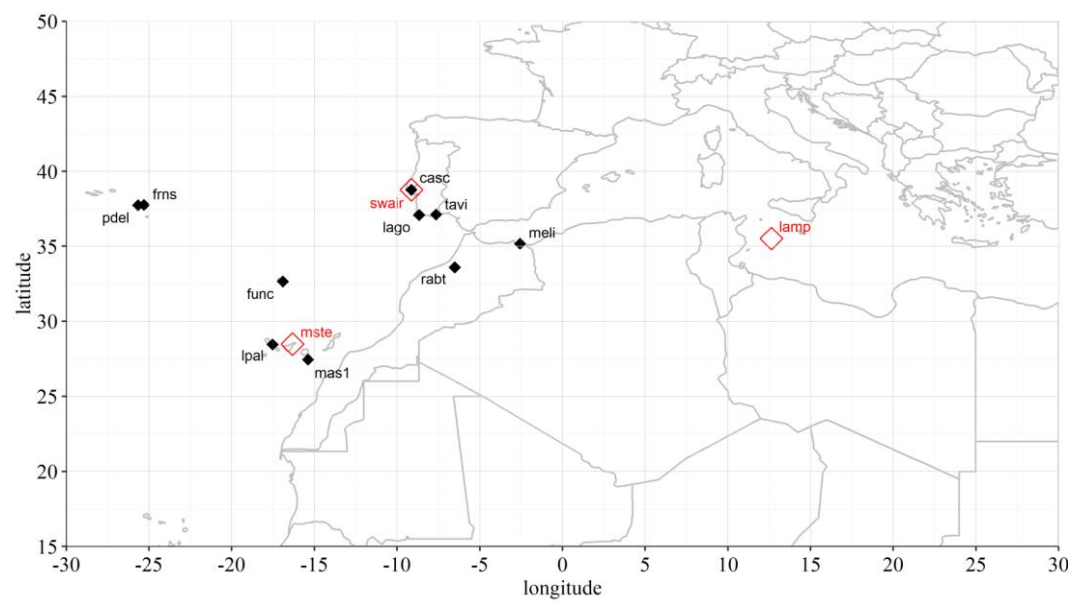

**Figure 1.** Locations of GNSS receivers used to calculate S4 (red open diamonds) and ROTI (black diamonds) scintillation indices.

**Lisbon.** The S4 data for Lisbon are obtained by a geodetic Septentrio GNSS receiver with SCINDA software installed in the area of the Lisbon airport (38.70° N, 9.14° W) between 2014 and 2019. The acquisition and processing of the data are described in detail in [21-22]. The data were validated during the PITHIA NRF TNA project ALERT [23].

**Lampedusa.** The S4 data for Lampedusa Island (35.52° N, 12.63° E) are obtained from a Novatel GSV4004B ISMR [24] who is operated at the Climate Observation Station of the Italian National Agency for New Technologies, Energy and Sustainable Economic Development (ENEA) between June 2011 and November 2018. The receiver owner is the INGV, and the data are available through the eSWua system (eswua.ingv.it, station code: "lam0s") [25]. The data on the slant S4 were used.

The GSV4004B was a single constellation (GPS only) scintillation receiver with multifrequency capabilities, able to evaluate amplitude scintillation by calculating the S4 index, which is the standard deviation of the received power normalized by its mean value [26]. Specifically, this is derived from the detrended received signal intensity. A high-pass filter is applied to detrend the raw amplitude measurements, as described by [24]. A fixed cutoff frequency of 0.1 Hz at 3 dB is used for both phase and amplitude filtering.

**Tenerife.** S4 data for Canary Islands are from a Trimble Zephyr Geodetic II receiver installed at Tenerife Island (28.48° N, 16.32°W) and operated by Universidad de La Laguna and the German Aerospace Center (DLR).

2.1.2 RINEX-based ROTI data

RINEX files from GNSS receivers installed in the studied area (black diamonds in Fig. 1) were used to calculate the ROTI index. Some of them belong to the Portuguese network of geodetic GNSS receivers RENEP (Rede Nacional de Estações Permanentes GNSS, https://renep.dgterritorio.gov.pt/) - pdel, frns, func, casc, lago, tavi. Other receivers are accessible through the FTP repository of the International GNSS Services (IGS: https://igs.org/).

*2.2 Japanese ROTI and TEC maps*

To study TEC and ROTI variations we also used products developed in the frame of the Dense Regional and Worldwide International GNSS-TEC observation (DRAWING-TEC) project led by Institute for Space-Earth Environmental Research (ISEE, https://aer-nc-web.nict.go.jp/GPS/DRAWING-TEC/). The two-dimensional maps we used (the absolute TEC and the rate of the TEC index, ROTI) are with the grid size of 0.5° x 0.5° and 5-min time resolution which we averaged to have 15-min time resolution. The maps are originally built using RINEX files obtained from more than 9300 GNSS receivers all over the world (as for January 2020).

*2.3 Swarm Ionospheric plasma bubble index*

Ionospheric bubble index (a Swarm product SW_IBIxTMS_2F, https://swarmhandbook.earth.esa.int/catalogue/sw_ibixtms_2f) was used to confirm EPB appearance in the studied area on June 22-23, 2015. The product provided flagged data related to the plasma bubble index.

In total, six events of EPB detection were found for the studied region and the studied time interval, however three of those EPB detections were far too East (at 18:30, 20:00 and 21:30 UTC of June 22,) and one EPB detection was far too West (June 23 02:00 UTC) from the studied region. Thus, only two EPB events detected by Swarm were used in this analysis: on June 22 at 23:00 UTC (Swarm A and C, latitudinal range 21°-30° N between 4.5° and 6° W in longitude) and on June 23 at 01:00 UTC (Swarm B, latitudinal range 27°-40° N and near 3° W in longitude).

*2.4 GIX, NeGIX and TEGIX indices*

This investigation also integrates a set of indices developed with the aim of monitoring and characterizing the perturbation degree of the ionosphere at mid-scales, in the order of few tens and up to some 200 km.

The gradient ionosphere index (GIX) [27] benefits from ground-based GNSS observations to estimate spatial gradients in the TEC domain. In this study, GIX values are computed using link related absolute/calibrated slant TEC instead of VTEC maps to minimize the ionosphere mapping errors. Ground-based data from the IGS network are used to create half hourly (with 1°x1° latitude and longitude resolution) maps of the 95-percentile metric of GIX, from 20:30 UTC of June 22 until 02:00 UTC of June 23.

Additionally, using Swarm satellites A and C, two data products have been recently developed to explore ionospheric perturbations at mid-scales, between 30 and 200 km [28]. The electron density gradient index (NeGIX, SW_NIX_TMS_2F) uses in-situ Langmuir Probe data to estimate plasma density gradients at the height of the spacecrafts, whereas the total electron content gradient index (TEGIX, SW_TIX_TMS_2F) uses level 2 GNSS data to compute TEC gradients in the topside ionosphere. Both products have a resolution of 0.5° in latitude, or about 8 seconds, along the track of the satellites. The near-parallel orbits of Swarm A and C, that reach a separation of ca. 180 km at the equator (in June 2015), and the combination of measurements within this resolution permits to compute gradient vectors not only meridionally along the path of the satellites, but also in the zonal (West-East) direction.

On June 22-23, 2015, Swarm satellites A and C observed the stronger phase of this geomagnetic storm above the Iberian Peninsula during their descending orbits, with an equatorial pass at 23h local time. Over the investigated region 65°W-65°E and 0°-50°N, five passes of the Swarm pair were recorded from the evening of June 22 to early hours of June 23: at 19:58 UTC, 21:32 UTC, 23:06 UTC, 00:40 UTC, and 02:14 UTC. Respectively, they flew along longitudes 41.5°E, 18°E, 5.5°W, 29°W and 52.5°W. A pass over the 50-degree latitudinal section lasts around 13 minutes.

## 3. Methods

*3.1 Multipath detection*

The multipath contamination of the S4 series was studied and the data for all three stations were cleaned using the two following approaches.

3.1.1 Threshold method

The first approach consisted of plotting the S4 data as a function of the ionosphere piercing points (IPP) date and time, as is shown in Fig. 2 for the Lisbon series. The diagonal lines in Fig. 2(a) correspond to individual satellites and result from the difference of the day length for a receiver (24 h) and revisiting time for each of the satellites (23h 56 min). Also, the plot shows a clear scintillation event (coloured vertical stripe) that took place during the night from June 22 to June 23. These diagonal lines show the multipath contamination, however, as one can see from Fig. 2(a), the values of S4 on these lines are not larger than 0.5-0.6. Thus, setting a threshold of S4 = 0.6 we can eliminate the multipath contamination as is shown in Fig. 2(b). As one can see, the threshold method removed the multipath signal, however, some data during the scintillation events of June 22-23 are also removed.

3.1.2 Sidereal S4 analysis

The sidereal multipath mask method relies on the repeated satellite visit every 23 hours and 56 minutes (i.e. sidereal day) of GPS satellites to detect the multipath sources (buildings, etc.) causing scintillation-like S4 index inflations [29-30]. Assuming the multipath sources are static for many days, when S4 is high for several days at the same sidereal time, that indicates a non-scintillation related signal disturbance. Scintillation is random by nature, and we do not expect to see the same S4 for consecutive days at the same exact sidereal time.

The average 7-day sidereal S4 (24 hours minus 4 minutes for GPS) is evaluated per PRN (pseudorandom noise codes unique to each of GNSS satellites). The maximal S4 value in these 7 days was excluded, i.e. considered an outlier, to avoid inflating the average sidereal S4 with the S4 resulting from scintillation. We use this average sidereal S4 as a mask to discard multipath contaminated data from the analysis. Figure 2(c) shows an example of the results of the sideral S4 analysis.

As one can see from the comparison of Figs. 2(b) and 2(c) (similar plots for Tenerife and Lampedusa can be found in the Supplementary Material, SM, Figs_S01_S02), the sideral S4 analysis provides a data set of S4 that is reliably cleared from the multipath effect and keeps the data related to the event. On the other hand, we found that the threshold method provides a more conservative version of a data set without a multipath effect and can be used for an initial data treatment, and the threshold should be selected after the analysis of each individual data set.

The S4 data cleaned by the sideral analysis were used further for the analysis of the June 22-23 event.

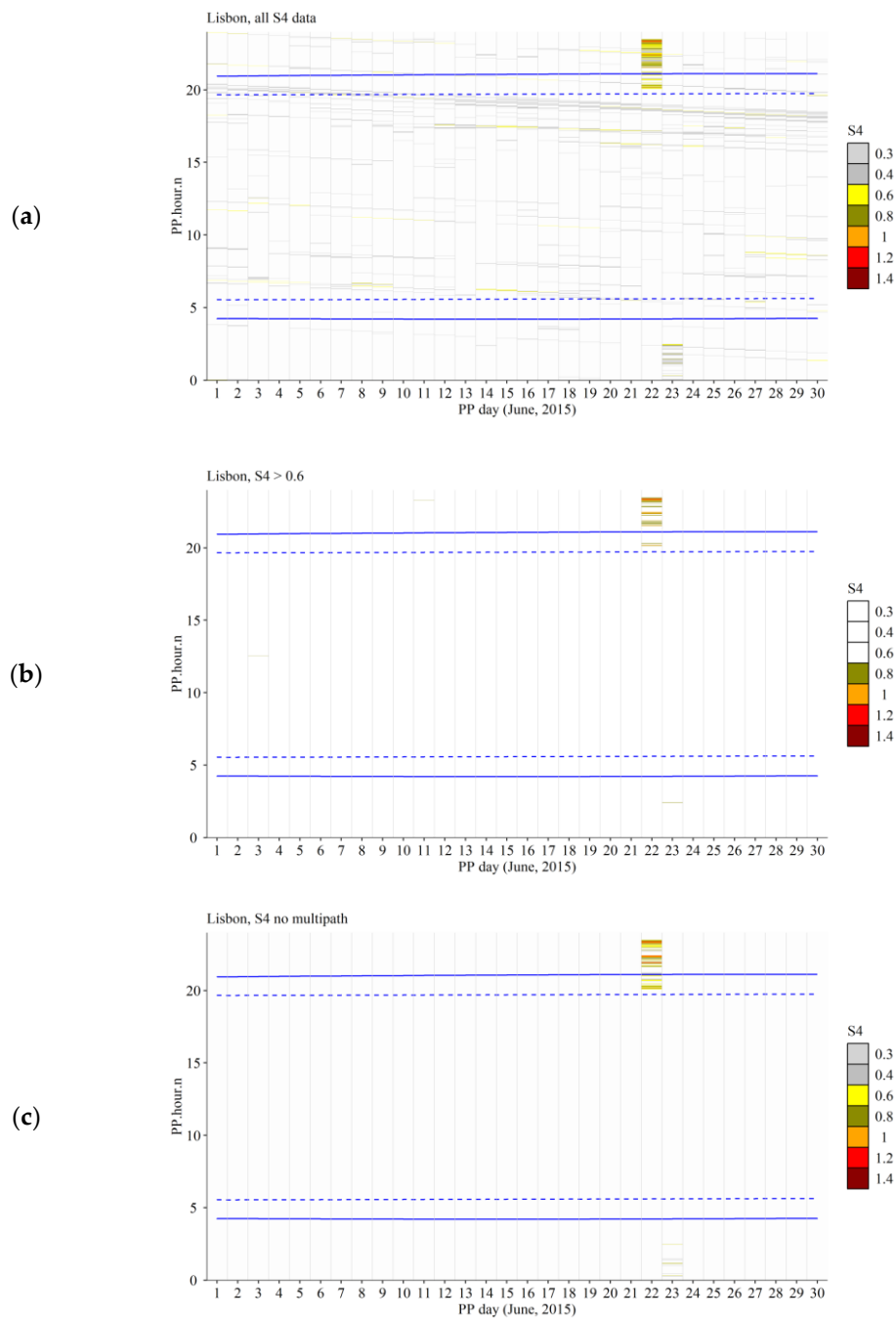

**Figure 2.** S4 data for Lisbon for June 2015. (a) All data. (b) Removing multipath contamination with an S4 threshold of 0.6. (c) Removing multipath contamination using sidereal S4 analysis.

*3.2 TEC gradients maps*

Spatial ionospheric gradients reflect spatial changes of the electron density and can be used to line out local areas with higher/lower electron density compared to the surrounding ionosphere. An analysis of the spatial TEC gradients can be used to detect boundaries of such structures as EPBs.

In this study we used the approach developed by [31] based on Global Ionospheric Maps (GIMs) to compute spatial gradients. Spatial TEC gradients are calculated as differences between TEC values at adjacent grid points (i,j) (see Fig. 3) at the same latitude ($\nabla xTEC$ or $\nabla x_j$ in Eq. 1) or longitude ($\nabla yTEC$ or $\nabla y$ in Eq. 2), which are then used to determine the absolute gradient ($\nabla TEC$ or $\nabla$ in Eq. 3):

$$\nabla x_{i,j} = (TEC_{i,j} - TEC_{i-1,j})/\Delta Lon, \qquad (1)$$

$$\nabla y_{i,j} = (TEC_{i,j} - TEC_{i,j-1})/\Delta Lat \qquad (2)$$

$$(\nabla_{i,j})^2 = (\nabla x_{i,j})^2 + (\nabla y_{i,j})^2 \qquad (3)$$

where $\Delta Lon$ and $\Delta Lat$ are grid steps (in km) along the longitude and latitude, respectively. In this work the spatial TEC gradients (in mTECu/km) were calculated for the studied region using TEC maps provided by the DRAWING-TEC project for the studied area between 25° W and 30° E and between 15° N and 50 ° N. The choice of the DRAWING-TEC products instead of GIMs is justified by the low spatial resolution (5° longitude by 2.5° latitude) of the openly available GIMs compared to the DRAWING-TEC products, as well as by significant smoothing of the TEC spatial variations resulting from the fitting of the GNSS data by the spherical harmonics basis functions during the GIMs production.

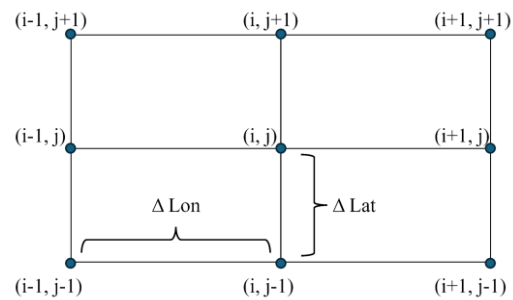

**Figure 3.** Scheme of the calculation of the spatial TEC gradients using TEC maps (adopted from [31]).

## 4. Scintillation event of June 22-23, 2015

*4.1 S4 from ground-based receivers*

The analysis of the June 22-23, 2015, scintillation event presented in this work is based on the data from three ground-based GNSS receivers located in the Mediterranean-Atlantic region (marked as open red diamonds in Fig. 1): in Lisbon and at Tenerife and Lampedusa islands. The Lisbon and Tenerife receivers are located close to the region(s) where anomalies in TEC and ROTI variations were detected in previous studies [1, 12], while the Lampedusa receiver is located well to the East from this region. Figure 4 shows S4 time variations during the days of June 22-23, 2015, observed at these three locations. As one can see, scintillations (S4 > 0.5) were observed at all three stations during the night June 22-23, 2015, starting from around 20h UTC, however the strength of the scintillation event depends on the location.

The highest S4 values were observed at Tenerife, also the peak of S4 was observed there earlier than at other stations (around 21h UTC). The time variations of S4 let us identify two periods of scintillations: on June 22 from ~19:30 to ~23:30 UTC (with maxima at ~21:00 and 21:45-22:00 UTC) and on June 23 from around midnight to ~03:00 UTC (maximum at ~01:00 UTC).

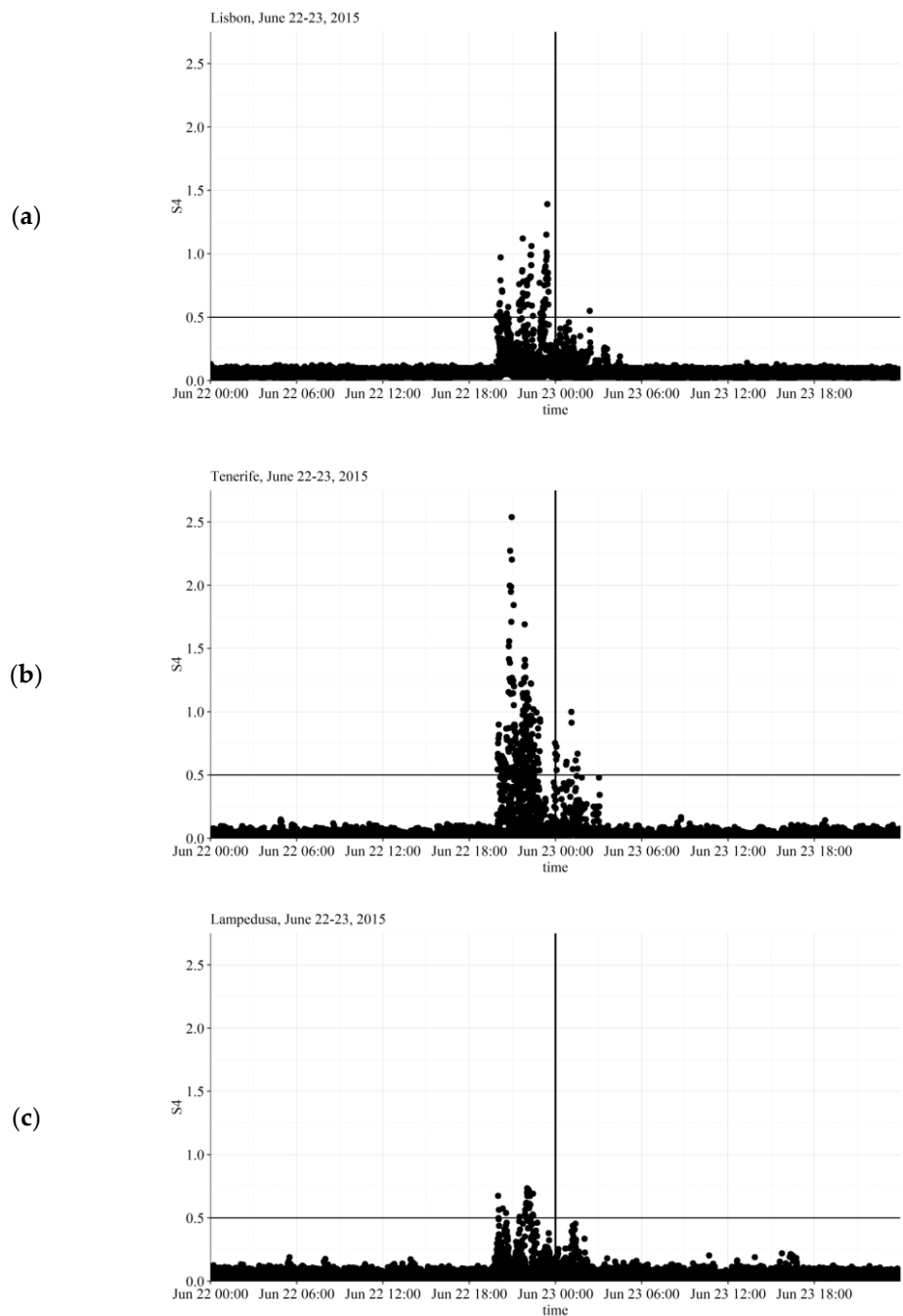

**Figure 4.** Time variations of S4 at Lisbon (a), Tenerife (b) and Lampedusa (c) during June 22-23, 2015. Vertical lines mark the midnight, horizontal lines mark S4 = 0.5. Time is in UTC.

Scintillations (S4>0.5) were observed at Lisbon, however the amplitude of S4 variations there is smaller than at Tenerife, and the peak of S4 was observed much later (near midnight). Also, at Lisbon it is possible to identify four periods of increased S4: on June 22 at ~20:00-21:00 UTC, ~21:30-22:30 UTC and, with the largest S4 values, at ~22:45-23:45 UTC, and on June 23 at ~02:00-02:30 UTC with S4 values near or slightly above the scintillation threshold.

While scintillations were observed by the Lampedusa receiver, their amplitude was much lower than at other two stations. This can be explained by the location of this receiver too far to the East from the scintillation sources. Here there were two periods of high S4 values on June 22 at ~20:00-20:30 UTC and ~21:30-22:40 UTC and two periods with S4 values below the threshold of S4 = 0.5 but still higher than the baseline: on June 22 at ~22:45-23:45 UTC and on June 23 at ~01:30-02:00 UTC.

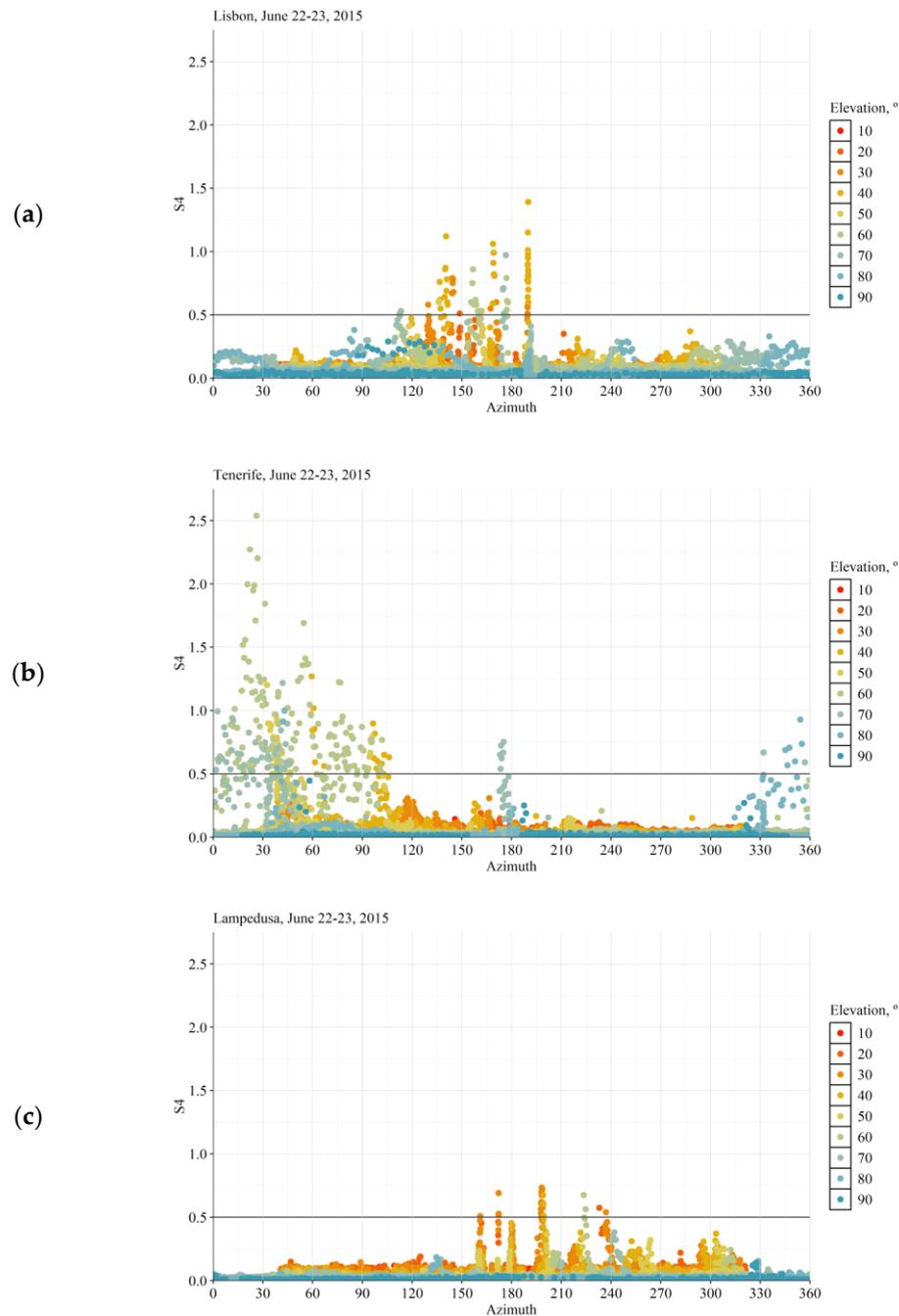

**Figure 5.** Distribution of S4 values (left axes) in dependence on the azimuth (X axes) and elevation (colours) of satellites at Lisbon (a), Tenerife (b) and Lampedusa (c). The horizontal lines mark S4 = 0.5.

Distributions of the S4 values depending on the azimuth and elevation of satellites during the days of June 22-23, 2015, are shown in Fig. 5 for all three stations. For Lisbon, all affected satellites were located to the South and South-East of the receiver (azimuth between 120° and 195°). Both low (< 20°) and high (≥ 20°) elevation satellites were affected. For Tenerife, most of the high S4 values were observed for the satellites located to the North and North-East (azimuth range 0°-105° and 330°-360°) from the receiver, also, S4 >

0.5 values were observed for satellites directly to the South of the receiver (azimuth 180°). All the satellites seen by the Tenerife receiver affected by the scintillations have elevations above 20° (some were at elevations close to 80-90°).

Satellites seen by the Lampedusa receiver that were affected by scintillations were in a broad range of azimuth (from ~150° to ~240°) and located in the South and South-West direction from the receiver. Almost all the affected satellites have elevation < 50°.

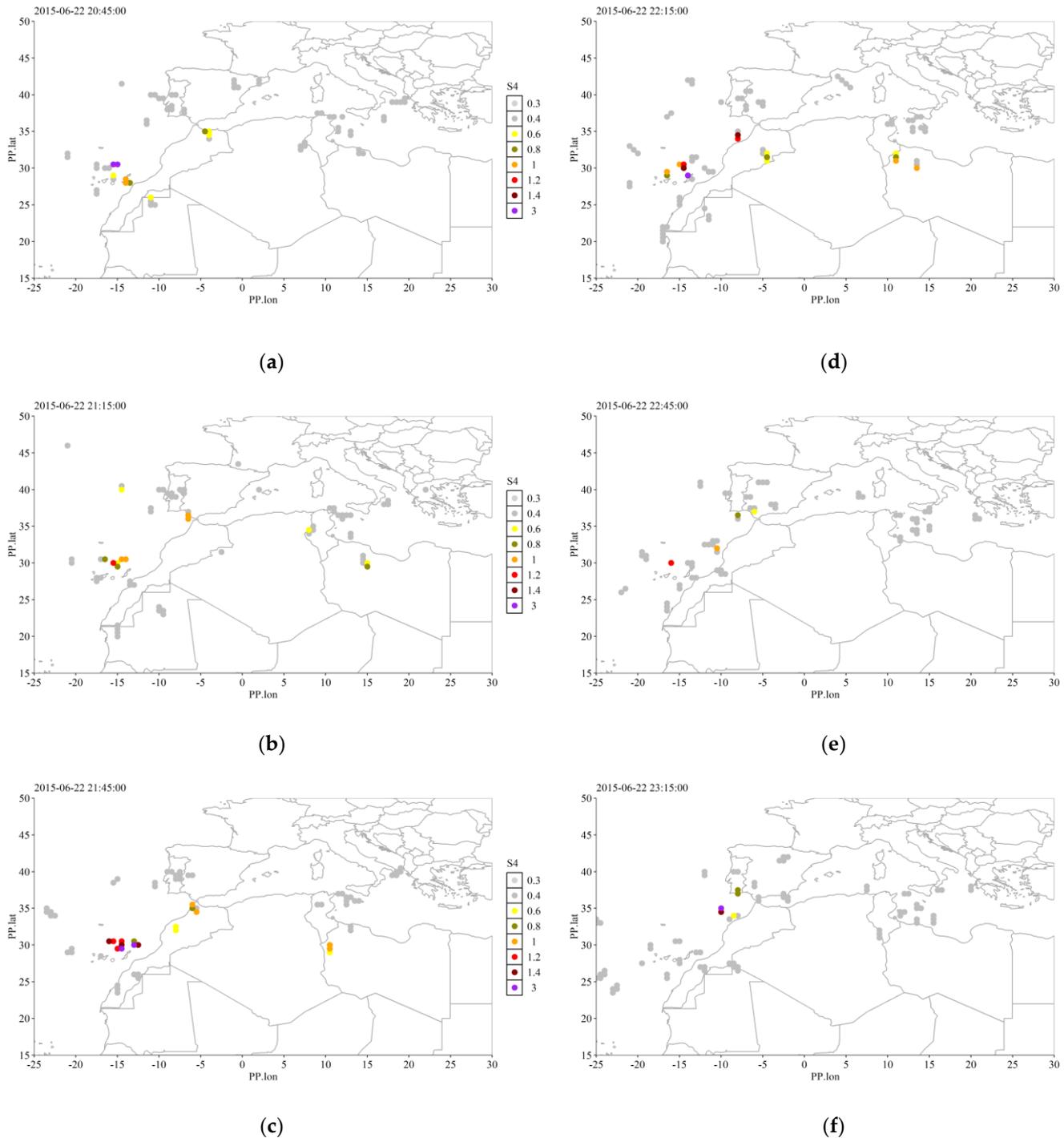

**Figure 6.** S4 (coloured dots) at IPP coordinates collected during 15 min on June 22 between 20:45 and 23:30.

The analysis of the azimuthal-elevation distribution of S4 for all three receivers allows us to assume that the sources of the scintillations observed in the studied region during the night of June 22-23, 2015, were travelling in the area between the Iberian Peninsula and the Canary Islands. This assumption is also confirmed by the maps shown in Figs. 6 and 7, and in SM (Anim_S01). These maps show S4 values at IPP (collected for 15 min time intervals starting from the timestamp of a map) on June 22 between 20:45 and 23:30 (Fig. 6) and on June 23 between 00:15 and 01:45 (Fig. 7). These time intervals are periods of the most intense scintillations (see Fig. 4).

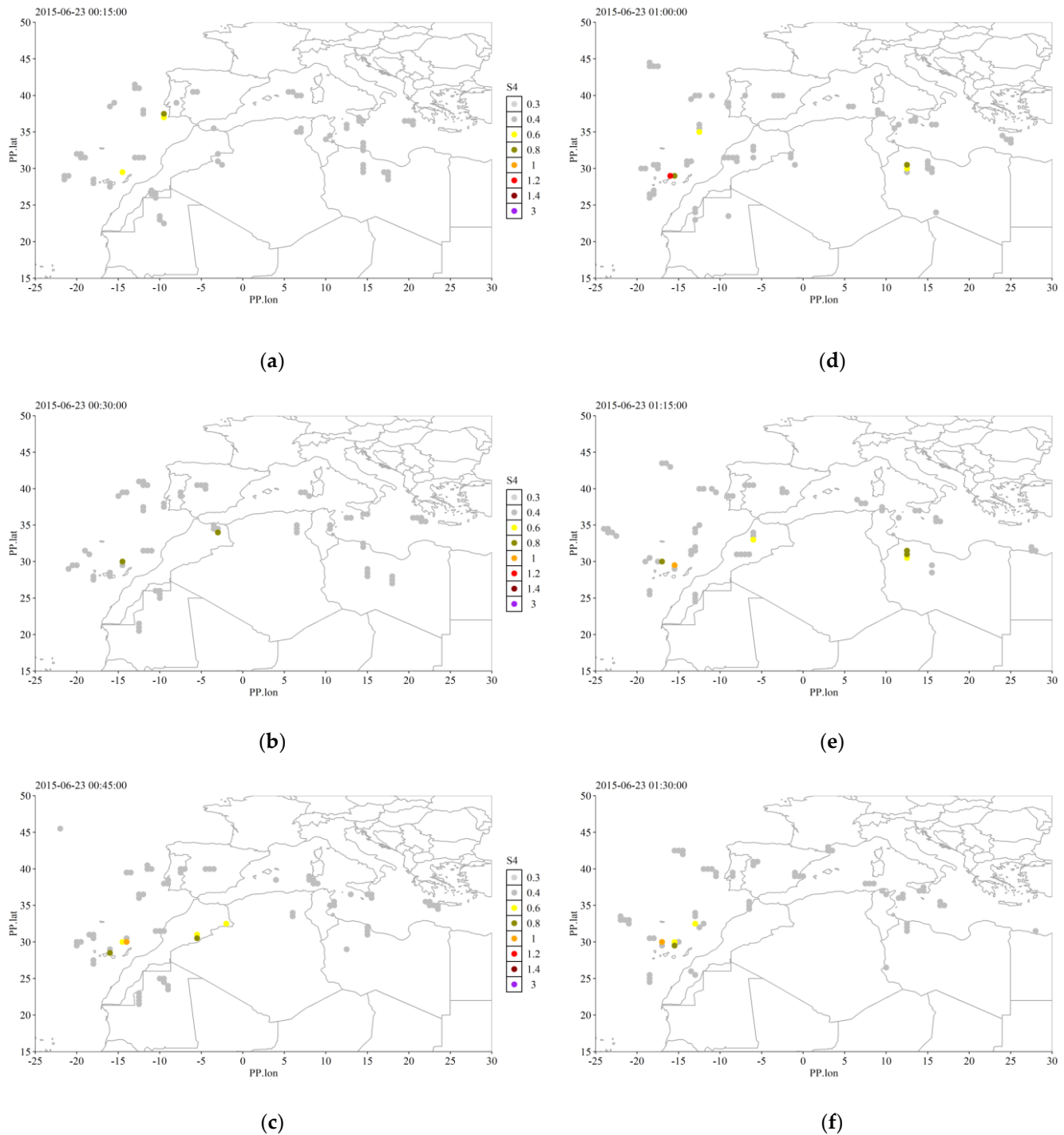

**Figure 7.** Same as Figure 6 but for June 23 between 00:15 and 01:45.

The maps show that IPP for the observations associated with high S4 values are mostly located in a region between the Canary Islands and the Iberian Peninsula, above North-Western Africa and the neighbouring areas of the Atlantic Ocean. Also, weak scintillations are associated with IPP over Northern Africa (to the South of Lampedusa).

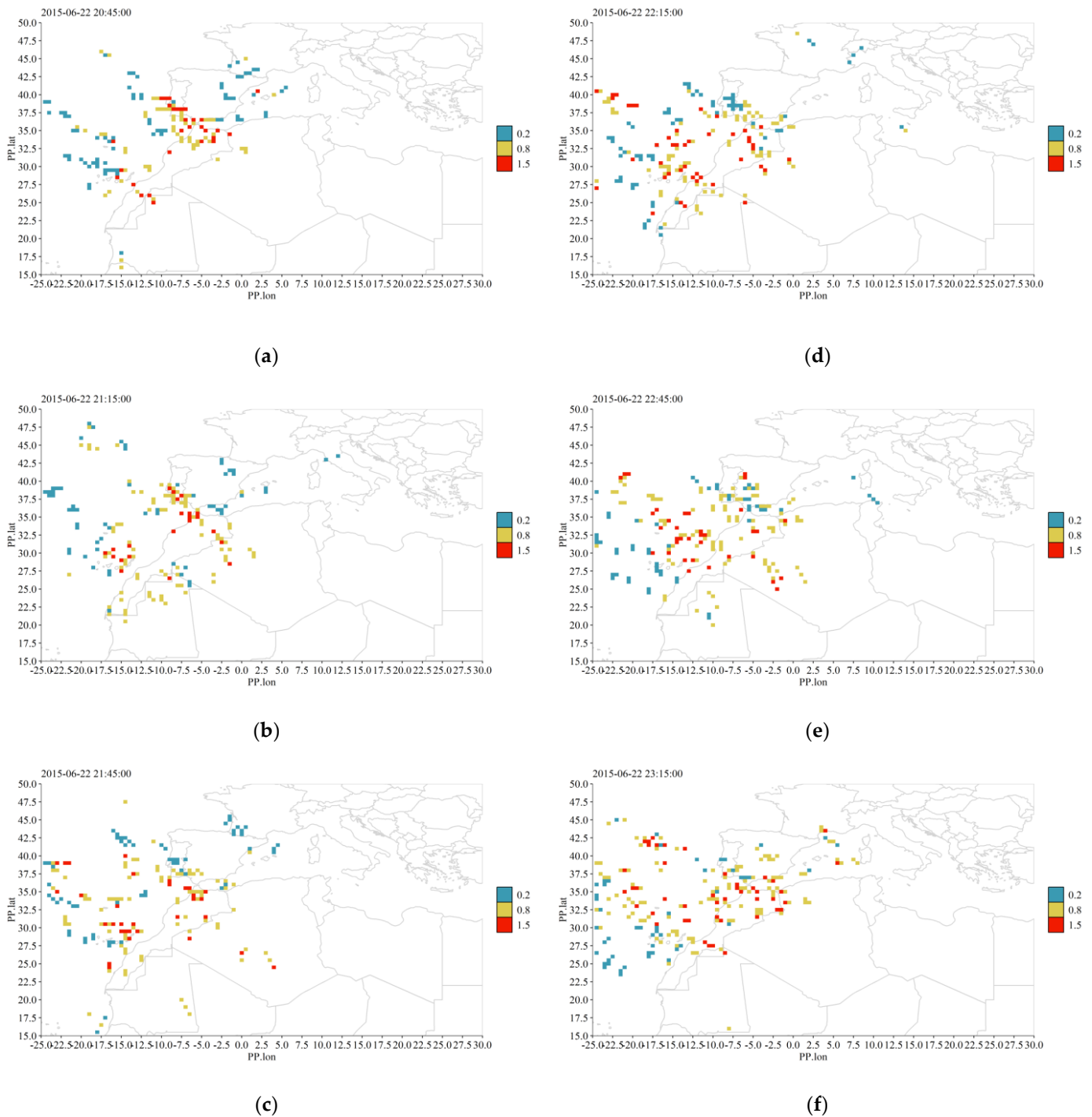

**Figure 8.** ROTI values from receivers (colour) binned to 0.5° × 0.5° grid collected during 15 min on June 22 between 20:45 and 23:30.

Based on the S4 data analysis, we conclude that the scintillations were caused by sources that passed over the Northern and North-Western Africa and further on over the North Atlantic Ocean. The scintillation sources were moving generally in the North-West direction as is seen in Anim_S01 (SM).

*4.2 ROTI indices: ground-based receivers and maps*

To study further the scintillation event of June 22-23, 2015, we analysed ROTI indices obtained both from the RINEX files of several receivers (black diamonds in Fig. 1) and from ROTI maps provided by the DRAWING-TEC project. Figures 8 and 9 show ROTI obtained from ground-based receivers in the studied area binned to the 0.5° x 0.5° grid for the same time intervals as Figs. 6 and 7 (see also SM Anim_S02).

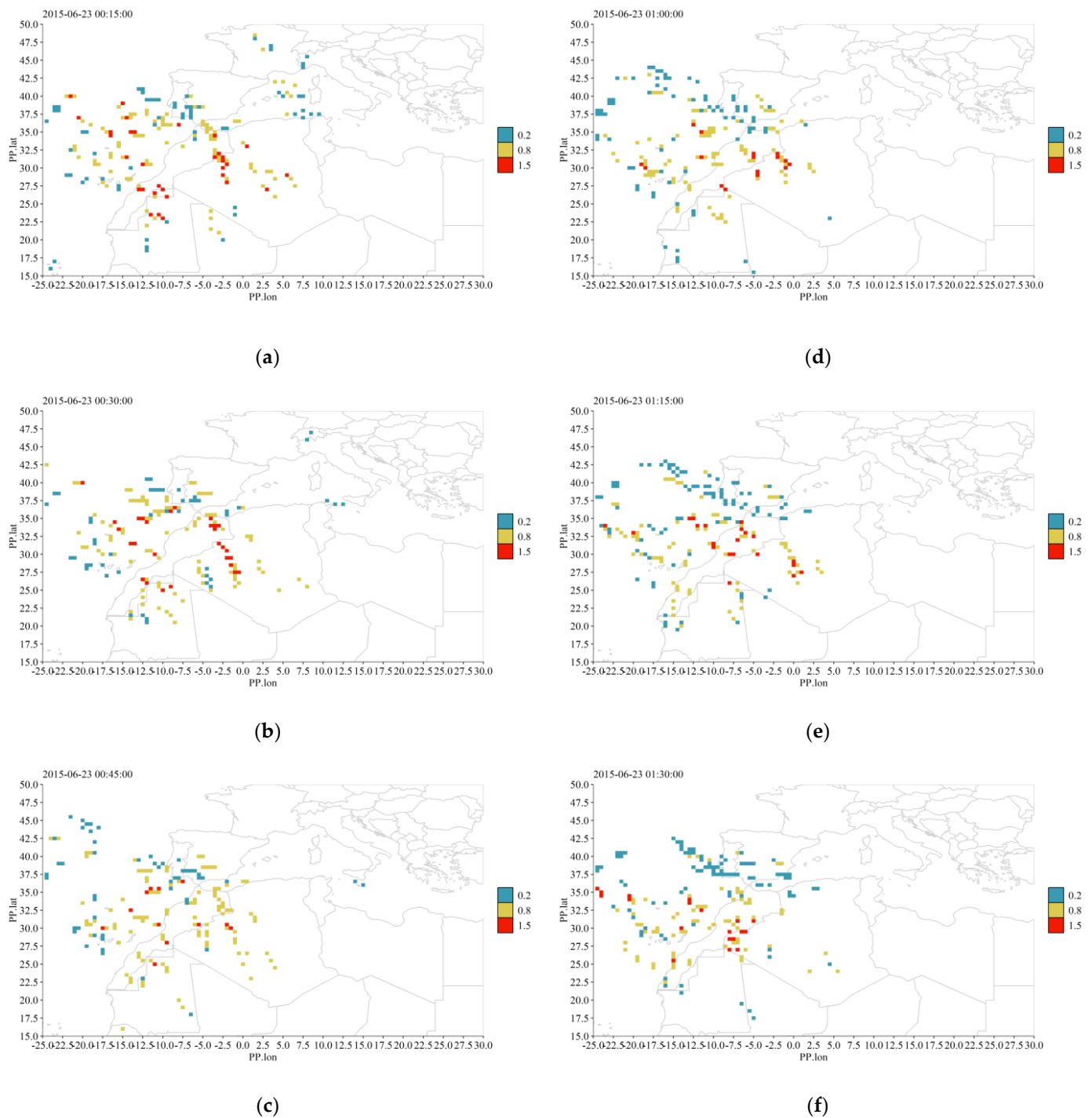

**Figure 9.** Same as Figure 8 but for June 23 between 00:15 and 01:45.

The ROTI data from the ground-based receivers confirm the finding based on the analysis of S4: the sources of scintillations during the studied time interval were in a wide band between the Canary Islands and the Iberian Peninsula. The strongest scintillations were observed in the late afternoon of June 22 (between 20:30 UTC and 23:30 UTC), followed by a smaller amplitude increase of the scintillations during the first hours of June 23 (between 00:00 UTC and 02:00 UTC). During the scintillation event the locations of the scintillation sources were slowly moving in the North-West direction.

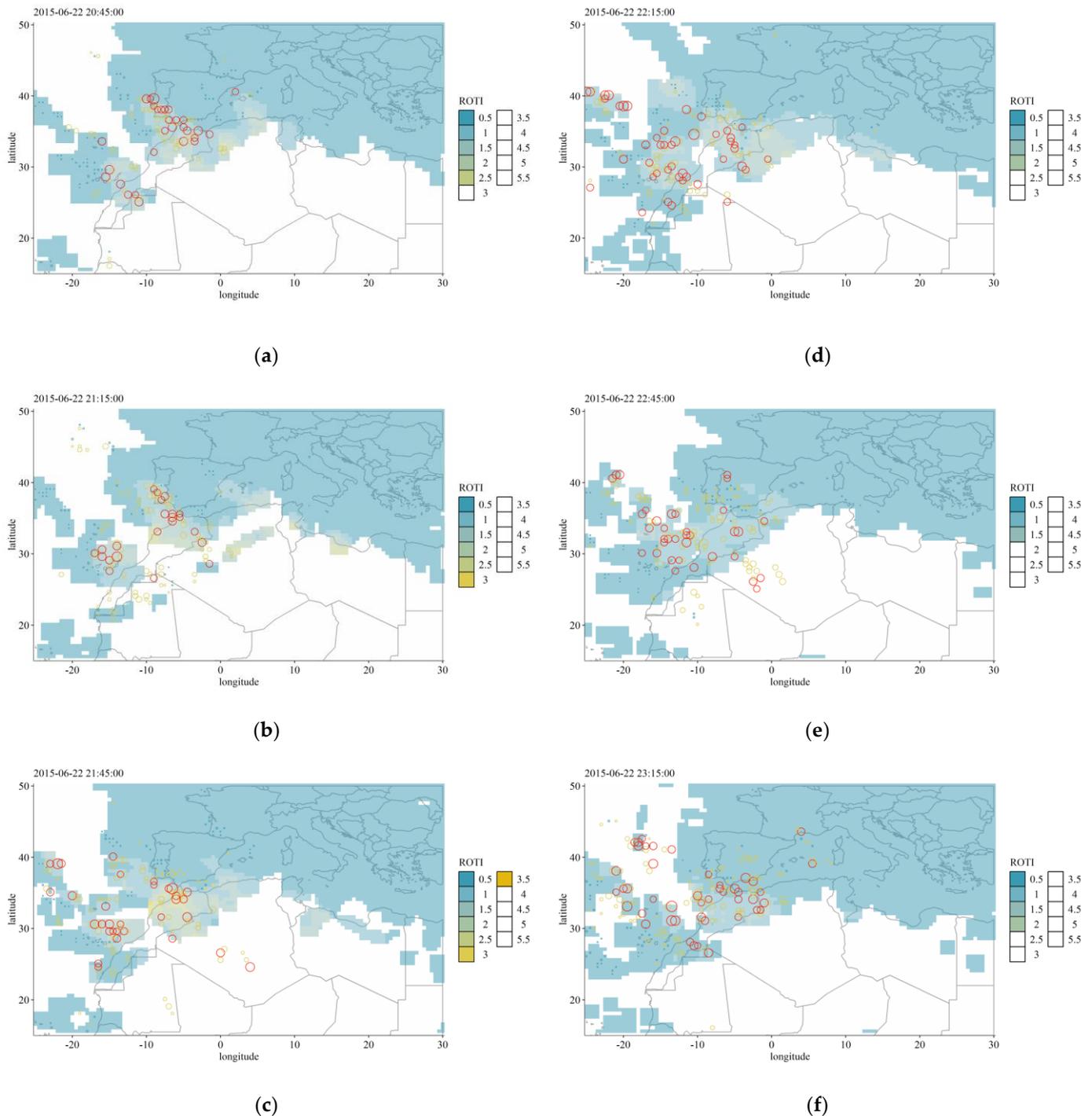

**Figure 10.** ROTI maps (colours) and ROTI data from the ground-based receivers (coloured open circles, see Figure 8 for the colour scheme, size is proportional to the ROTI values) on June 22 between 20:45 and 23:30.

This conclusion is confirmed by the ROTI. Figures 10 and 11 show ROTI maps for the same time intervals as in Figs. 8 and 9 (colour scheme) with superimposed ROTI data from the ground-based receivers (coloured open circles) and the Swarm ionospheric plasma bubble index (grey dots along meridians). An animation can be found in SM (Anim_S03).

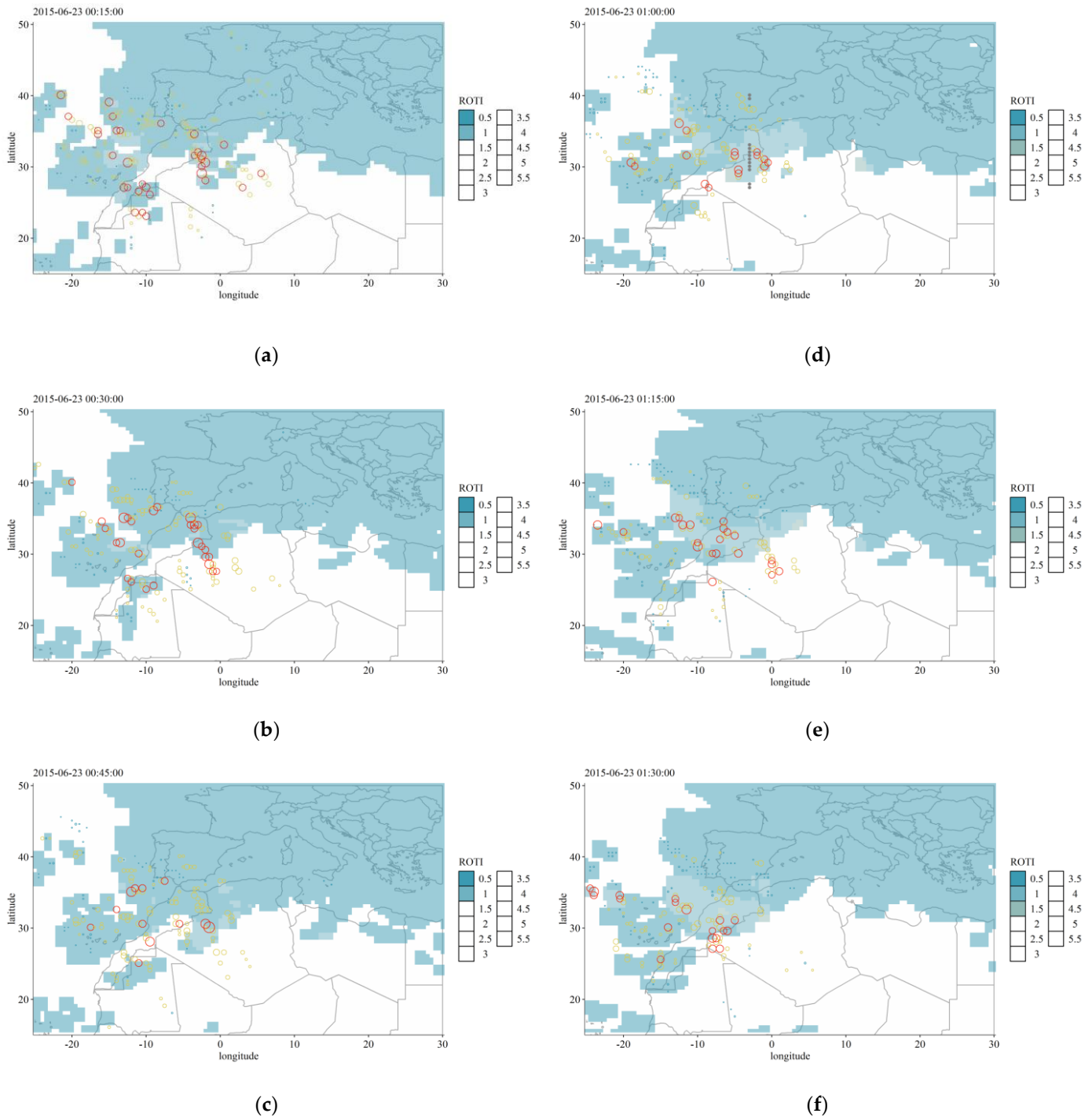

**Figure 11.** Same as Figure 10 but for June 23 between 00:15 and 01:45. Swarm ionospheric plasma bubble index data are shown as grey dots along meridians in (d).

Since S4 and ROTI indices can be considered as measures to detect the electron density inhomogeneity on different spatial scales [19] - smaller scales (few hundreds of meters) for S4 and medium scales (few kilometres) for 5-minute ROTI from 30 s RINEX at low-latitudes - we compared the ROTI and S4 data. Figures 12 and 13 show ROTI maps for the same time intervals as before (colour scheme) with superimposed S4 data from the ground-based receivers (white/black open circles, see figure captions) and the Swarm ionospheric plasma bubble index (grey dots along meridians). An animation can be found in the SM (Anim_S04).

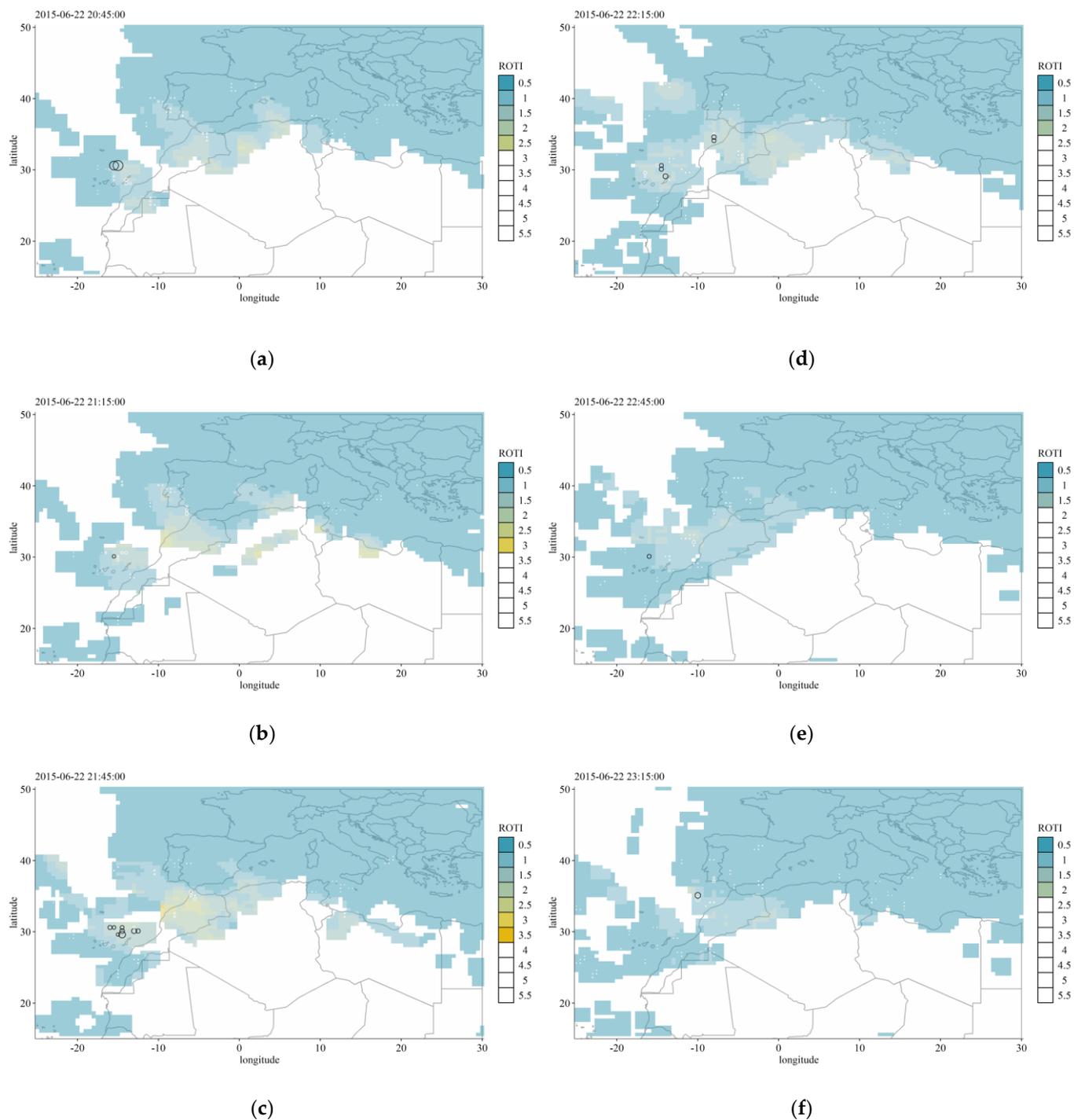

**Figure 12.** ROTI maps (colour) and S4 measurements (open circles: white for S4 < 0.8 and black for S4 ≥ 0.8, size is proportional to S4 values) on June 22 between 20:45 and 23:30.

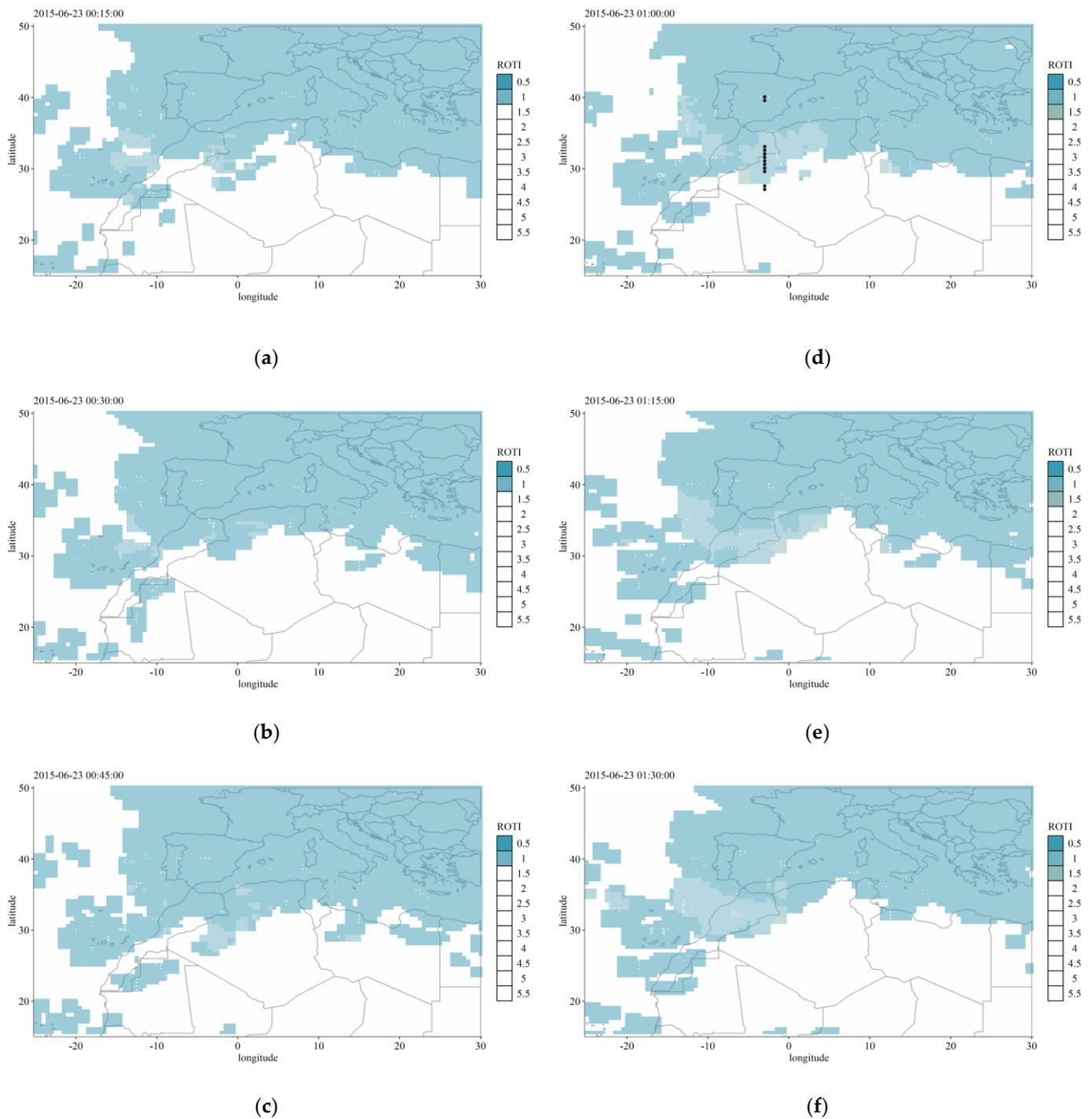

**Figure 13.** Same as Figure 12 but for June 23 between 00:15 and 01:45. Swarm ionospheric plasma bubble index data are shown as grey dots along meridians in (d).

The S4 data are available only for three ground-based receivers and are limited to the satellites available during the studied time interval. Thus, the spatial coverage of the S4 data is much sparser than of the ROTI data. However, high S4 values (black circles) tend to cluster in the regions with higher ROTI values (green-to-reddish colours). The link between ROTI and S4 is more clearly seen during the night hours of June 22 than during the early hours of June 23 since the amplitudes of both S4 and ROTI variations during the scintillation event were much higher on June 22.

The comparison of the S4 and ROTI variations allows us to conclude that the scintillation event of June 22-23, 2015, was caused by inhomogeneities in the ionospheric electron density that existed at different spatial scales.

*4.3 TEC gradients and plasma bubbles identification*

Ionospheric plasma bubbles (EPBs) are ionospheric inhomogeneities elongated in the meridional and altitudinal directions containing plasma with electron density much lower than the surrounding ionosphere. Large EPB [3, 32] may have a complex internal structure with regions of higher and lower electron density.

EPBs were previously reported as main sources of the ionospheric scintillations observed in the studied region during the night June 22-23 [1, 12]. To identify these structures, we used TEC spatial gradients calculated using TEC maps provided by the DRAWING-TEC project. Spatial (E-W and N-S) gradients of TEC and their variability have been demonstrated to be very effective in proxying the related scintillation on GNSS signals due to EPBs [33]. The maps of the absolute TEC gradients are shown in Figs. 14-15 and 16-22 for the selected time intervals on June 22 and 23 with overplotted ROTI and S4 data, respectively. Animations containing similar plots for other time intervals as well as the data for the longitudinal ($\nabla x TEC$) and latitudinal ($\nabla y TEC$) gradients can be found in SM (absolute gradients and ROTI: Anim_S05, longitudinal gradients and ROTI: Anim_S06, latitudinal gradients and ROTI: Anim_S07; absolute gradients and S4: Anim_S08, longitudinal gradients and S4: Anim_S09, latitudinal gradients and S4: Anim_S10).

As one can see, the areas of high ROTI and S4 values coincide with regions of high spatial TEC gradients (Figs. 14-22). Spatiotemporal patterns of the variations of the scintillation indices and TEC gradients are very well correlated. These patterns include, for example, several structures stretched along the North-West - South-East direction that are seen both in the maps of the spatial gradients and of ROTI and are located between the Canary Islands/Madeira and the Iberian Peninsula. Such structures are not clearly visible in Figs. 6 and 7 which show S4 values at corresponding IPP, however, when the S4 values are overplotted on the TEC gradients maps, it is seen that most of IPP of the signals with the high S4 values are located inside these structures. While these elongated structures were previously reported for ROTI [1, 12], in this work, by comparing the data for the spatial gradients, and the ROTI and S4 data, we can conclude that the high TEC gradients depict boundaries of inhomogeneities in the ionospheric electron density that caused ionospheric scintillations.

Such relations between the scintillations and the spatial gradients were already shown, for example, in the climatological analysis for the Brazilian region [33]. In this work we confirm this relation for a different region and for an individual scintillation event. Also, we can confirm another result of [33]: in general, the latitudinal ($\nabla y TEC$) TEC gradients are higher than the longitudinal ($\nabla x TEC$) ones, and, consequently, the contribution of the latitudinal gradients to the values of the absolute TEC gradients is also larger in general.

Figures 15(d) and 22(d) show EPB identified by the Swarm mission on June 23, 2015, 01:00 UTC (vertical lines of grey dots depict Swarm PBI = 1). They are shown again in Fig. 18 together with similar plots for June 22, 2015, 23:00 UTC. For the plasma bubble identified by the Swarm product at 23:00 UTC on June 22 as containing EPB (vertical lines of grey dots) coincide with areas of high absolute (see Fig. 18(a-b)) and longitudinal (see SM Anim_S09 and Anim_S10) TEC gradients, but not with the areas of high ROTI values of the ROTI maps (see Fig. 18(c)). On June 23 at 01:00 UTC the areas identified by the Swarm product as containing EPB are located between the areas of high TEC gradients (Fig. 18(d-e))/high ROTI values of the ROTI maps (Fig. 18(d and f)). One of the possible explanations for no relations between the spatial TEC gradients and EPB identified by Swarm at this time is that the structures identified by the Swarm product have smaller size than the TEC maps resolution (0.5° x 0.5°).

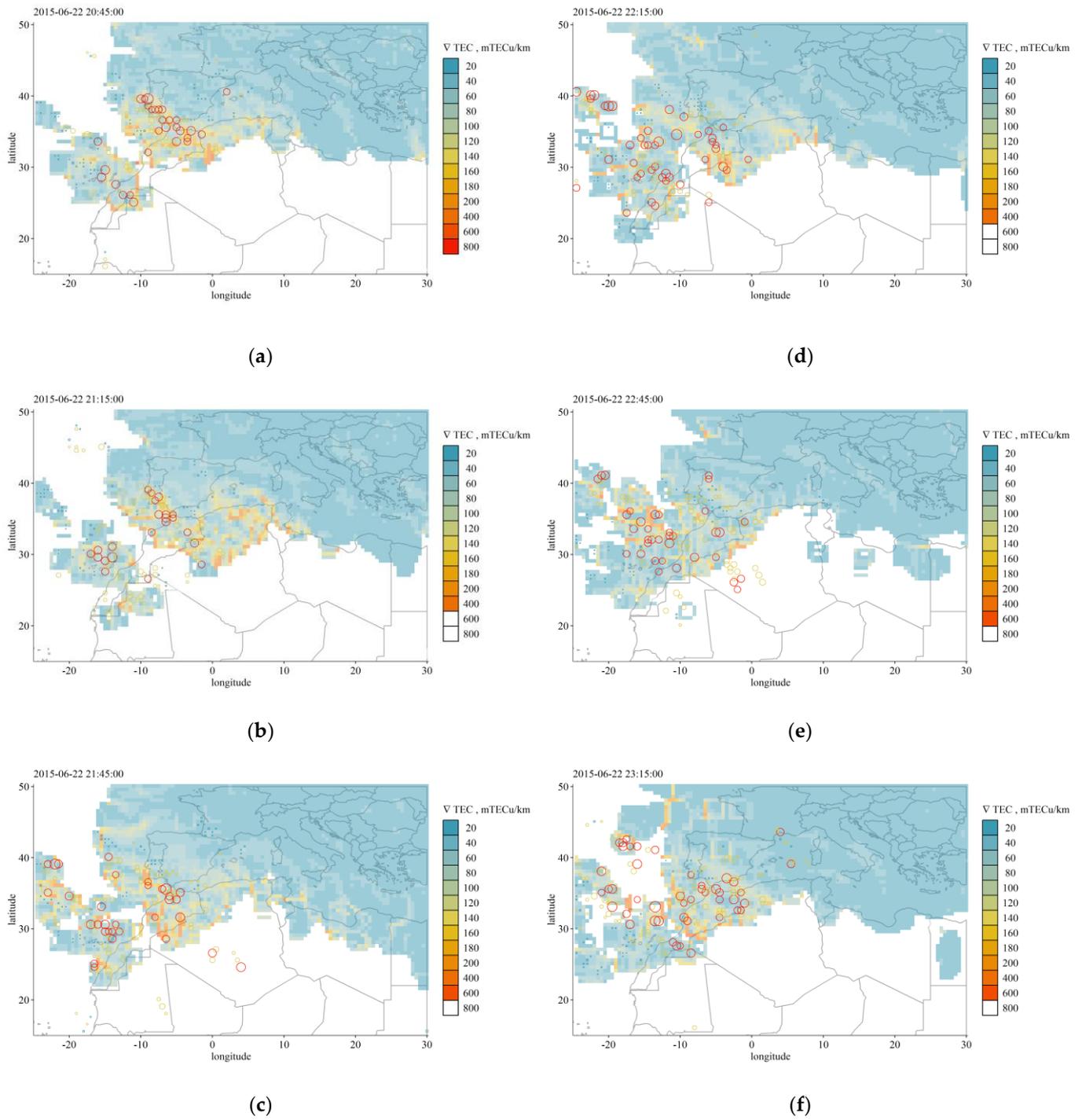

**Figure 14.** Absolute TEC gradients (colour) calculated for 15 min time intervals on June 22 between 20:45 and 23:30. ROTI data from receivers are shown as open circles (coloured open circles, see Figure 8 for the colour scheme, size is proportional to the ROTI values).

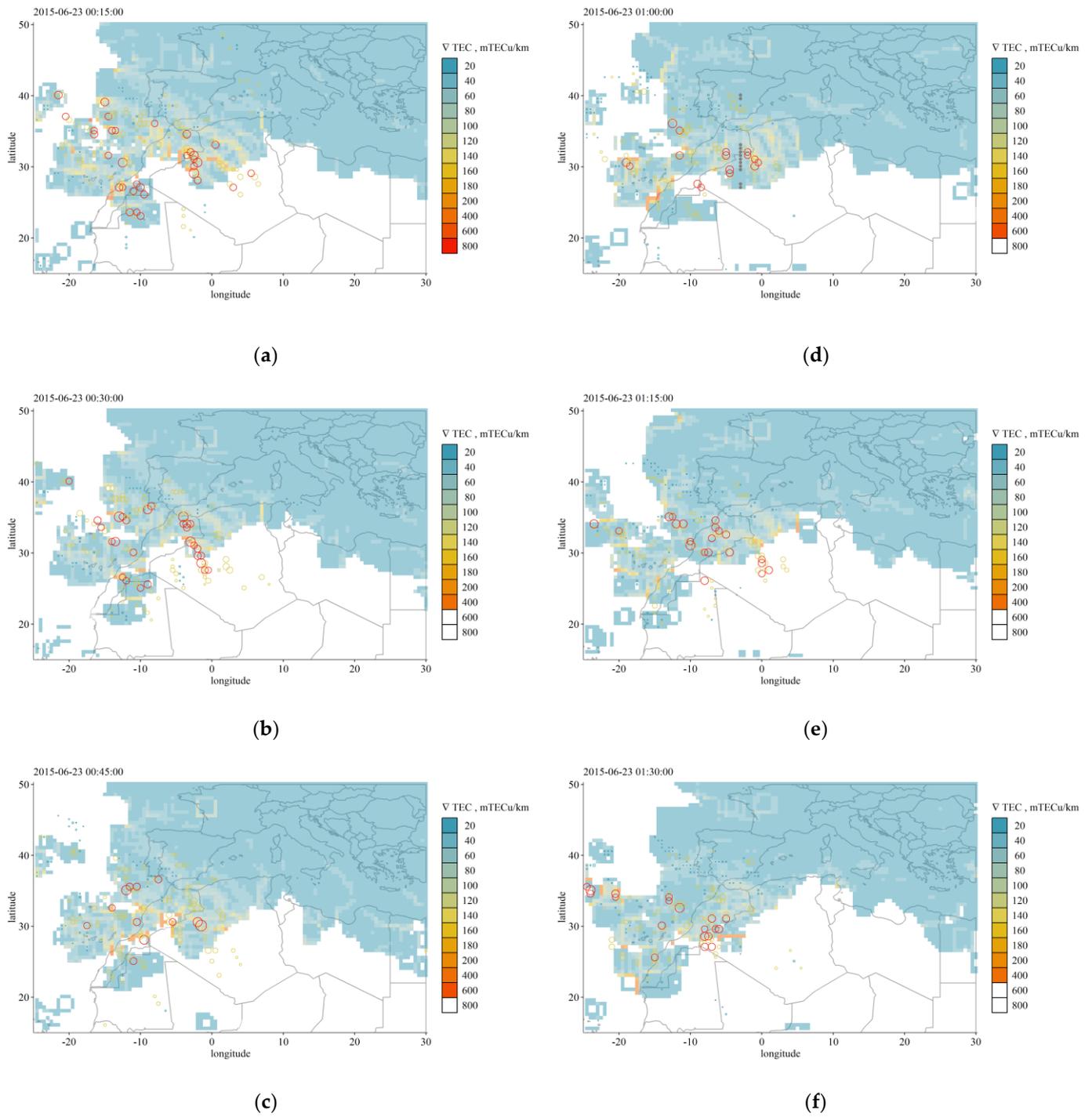

**Figure 15.** Same as Figure 14 but for June 23 between 00:15 and 01:45. Swarm ionospheric plasma bubble index data are shown as grey dots along meridians in (d).

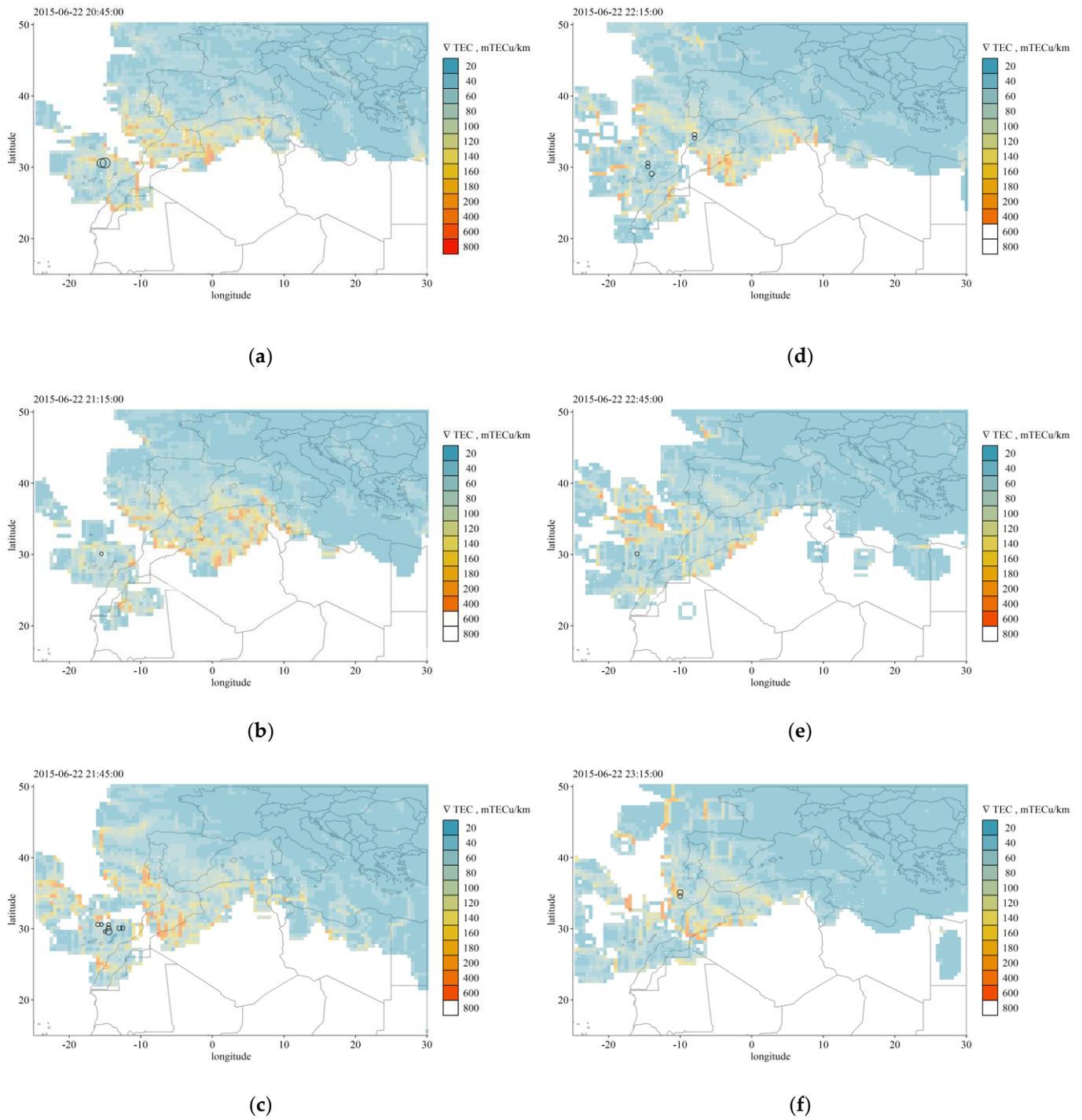

**Figure 16.** TEC gradients (colour) calculated for 15 min time intervals on June 22 between 20:45 and 23:30. S4 measurements are shown as open circles (see Figure 12 for the colour scheme, size is proportional to S4 values).

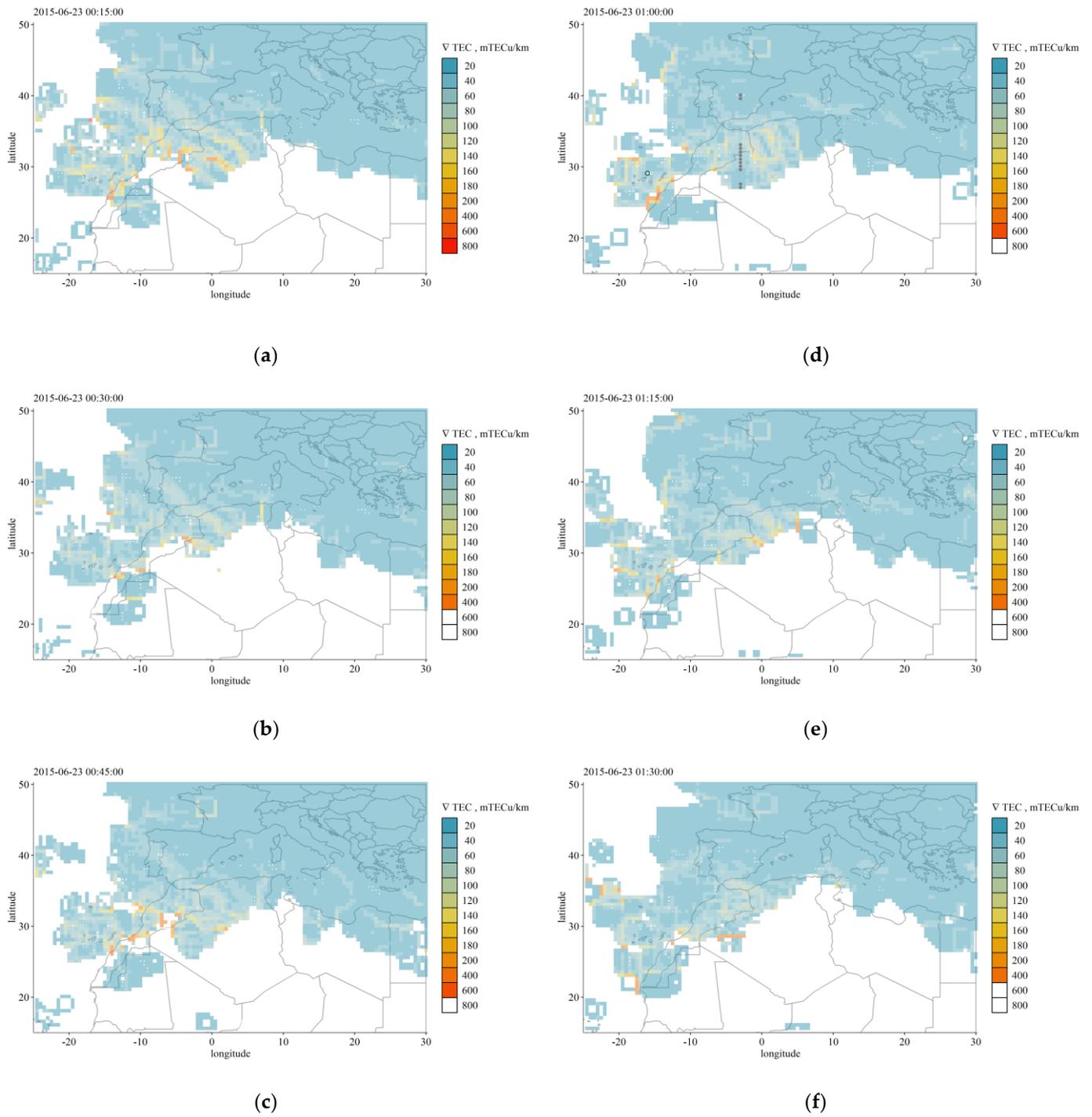

**Figure 22.** Same as Figure 16 but for June 23 between 00:15 and 01:45. Swarm ionospheric plasma bubble index data are shown as grey dots along meridians in (d).

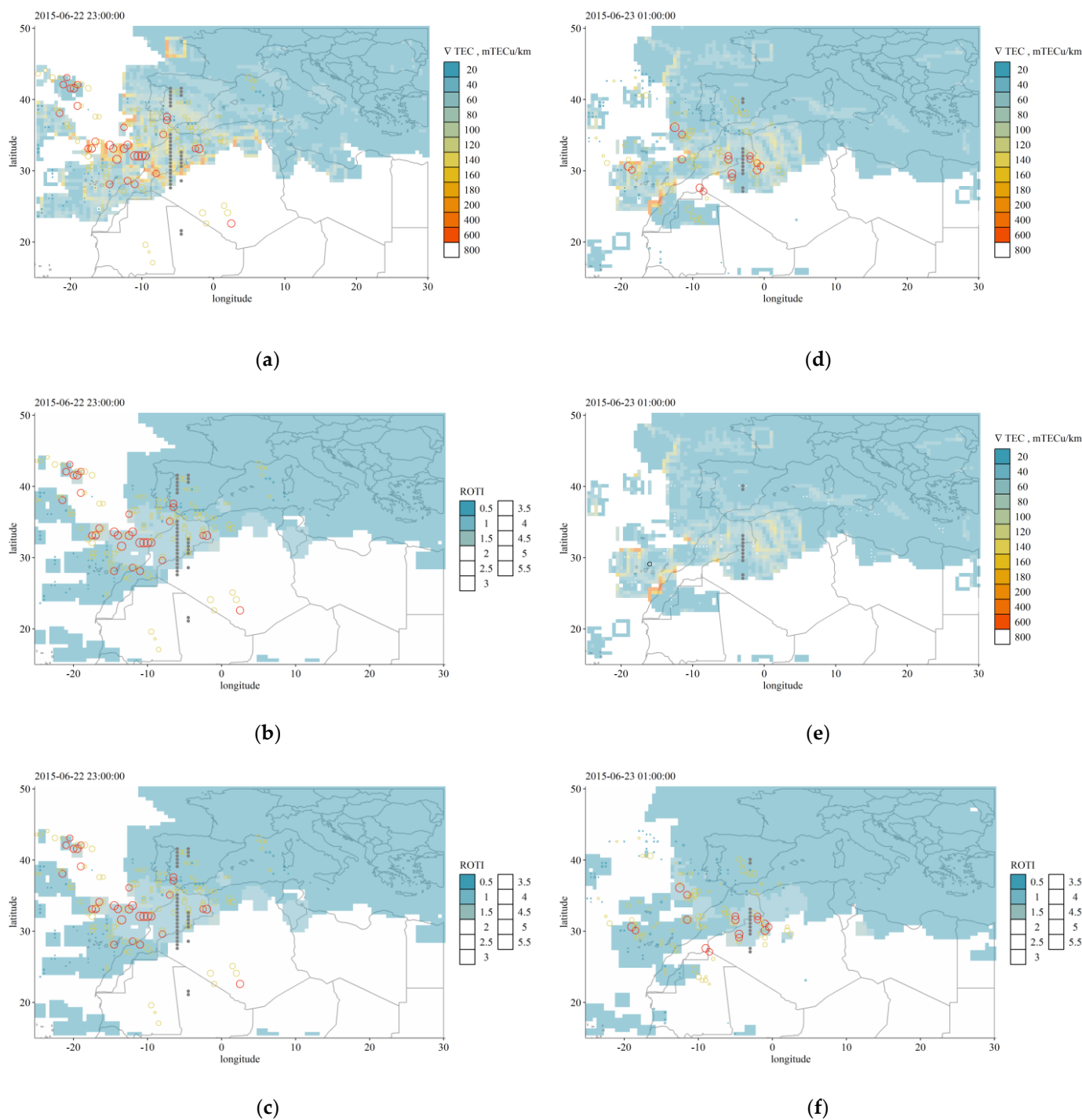

**Figure 18.** Maps of the absolute TEC gradients (a-b and d-e, colours) and ROTI (c and f, colours) together with scintillation indices obtained from ground-based receivers: ROTI (a, c, d, f, coloured circles, see Figure 8 for the colour scheme, size is proportional to the ROTI values) and S4 (b, e, white and black circles, see Figure 12 for the colour scheme, size is proportional to S4 values), and the Swarm PBI detections (vertical lines of grey dots) for June 22 23:00 and June 23 01:00 UTC.

There are other ways to confirm relations between ionospheric scintillations and depletions of the electron density, revealing the presence of EPBs. For example, by the analysis of slant TEC (sTEC) variations obtained for satellites experiencing signal scintillations [18]. To perform this analysis we selected four satellites, both for the Lisbon and Tenerife receivers, with the highest S4 values observed during the night June 22-23. These are satellites with PRN G09, G19, G23 and G32 for Lisbon and G01, G04, G11 and G32 for Tenerife.

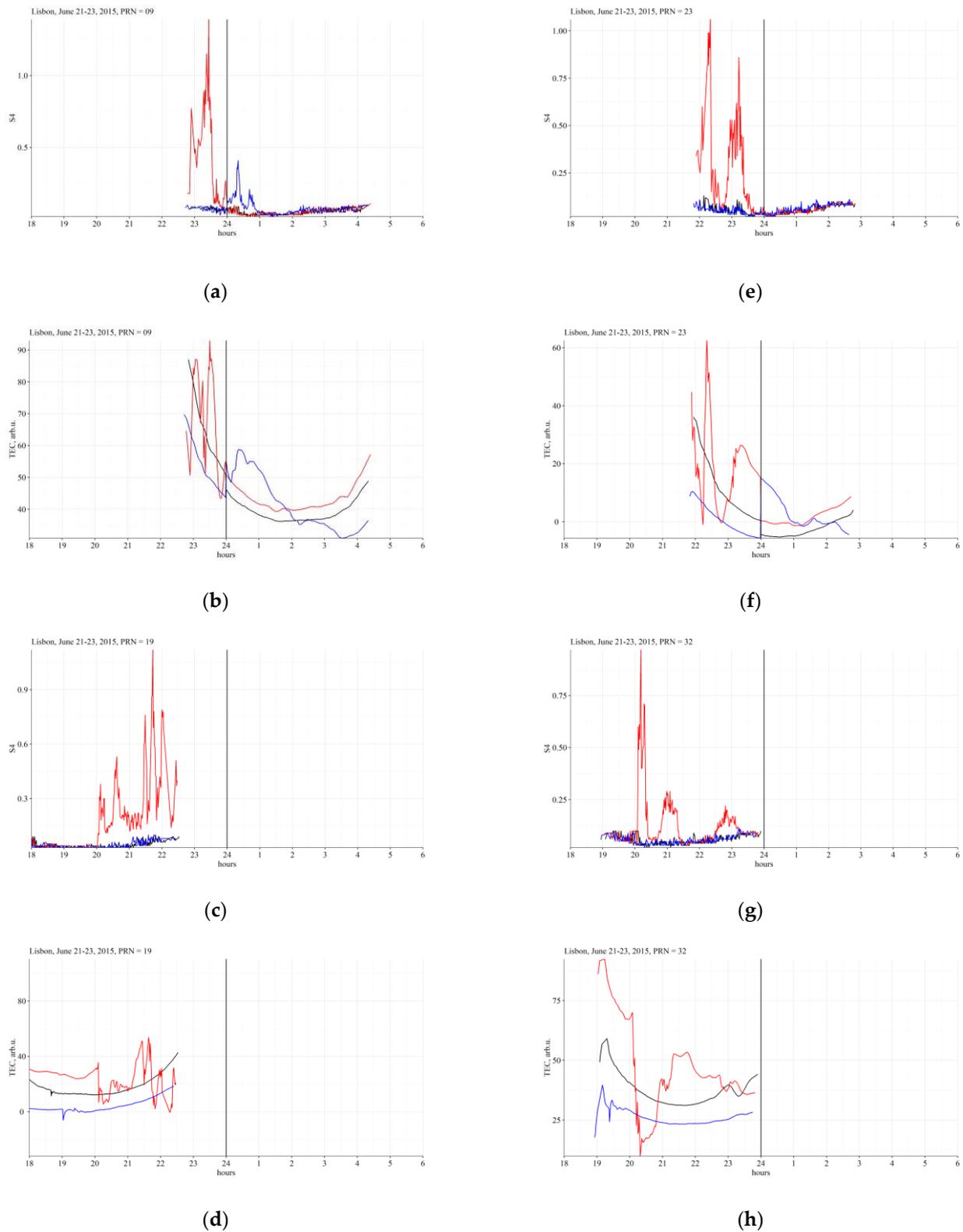

**Figure 19.** S4 (a, c, e, g) and sTEC (b, d, f, h) values obtained at Lisbon for PRNs G09 (a-b), G19 (c-d), G23 (e-f) and G32 (g-h) between 18:00 UTC and 06:00 UTC of the following day for June 21-23, 2015. Colours: June 21 - black, June 22 - red, June 23 - blue.

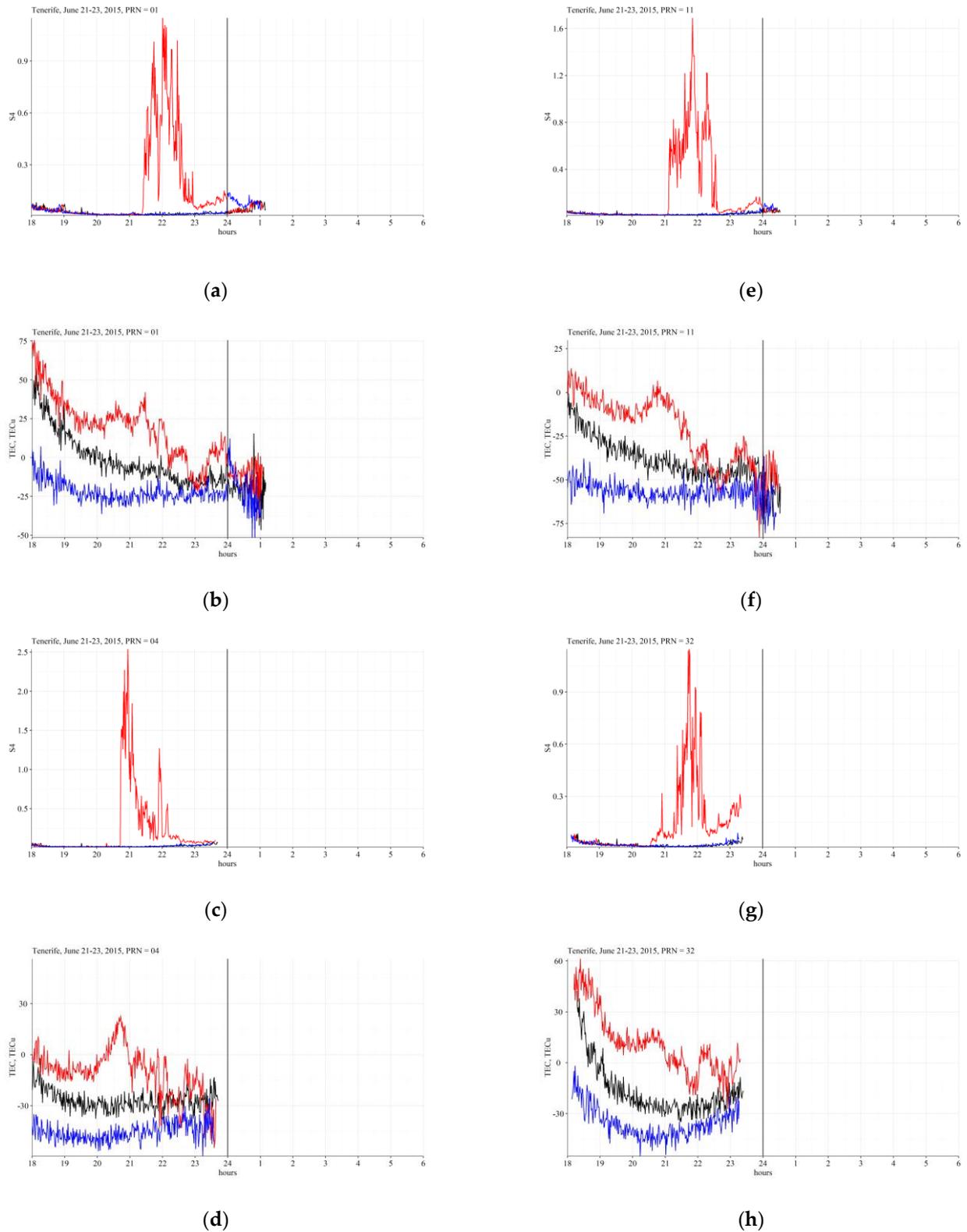

**Figure 20.** Same as Figure 19 but for Tenerife and PRNs G01 (a-b), G04 (c-d), G11 (e-f) and G32 (g-h).

Figures 19 and 20 show variations of S4 and sTEC during the studied time interval for these satellites. Please note that sTEC values for Lisbon are in arbitrary units [see 21-22]. As one can see, after the geomagnetic storm commencement in the late afternoon of June 22, 2015, sTEC values began to rise following the positive ionospheric storm development (as described, e.g., in 10, 11, 7, 9). However, when the S4 values started to grow around 20-22 h UTC (depending on the receiver and PRN), the sTEC values dropped significantly (by 30-50 TECu as measured at Tenerife). The times of the scintillations' onsets and the sTEC decreases are well correlated for each of the satellites. According to [18], this S4-sTEC dynamics can be interpreted as an effect of a() EPB(s) crossing the line of sight (LOS) of a receiver-satellite pair. Figure 21 shows IPP of the same satellites as in Figs. 19 and 20 during the time intervals of the strongest scintillations for each of the satellites. As one can see, the directions of LOS (and locations of corresponding sTEC decreases related to EPBs) coincide very well with areas of high spatial TEC gradients (see Figs. 16 and 18) allowing us to conclude that these high spatial gradients result from EPBs crossing the studied area.

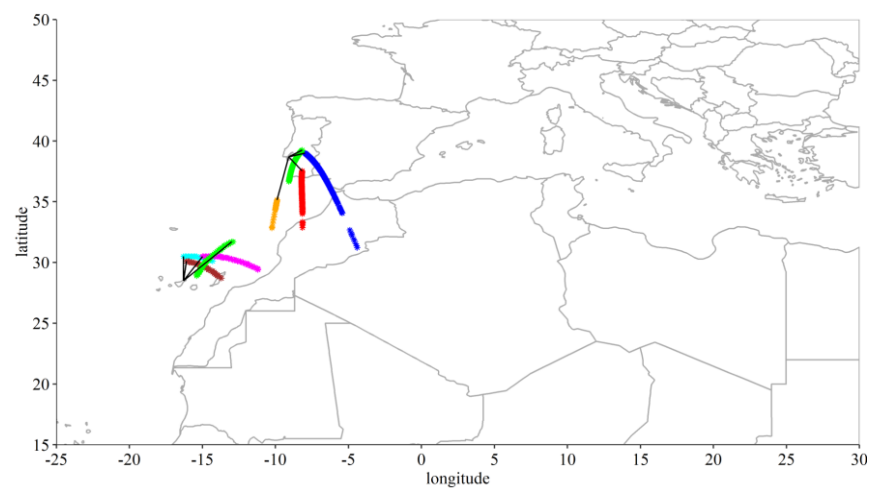

**Figure 21.** IPP of several GPS satellites (see text for details) - coloured asterisks, observed at Lisbon and Tenerife during the strong scintillations (slightly different time intervals for different satellites). Black lines connect IPP trajectories to corresponding GNSS receivers and give a simplistic visualization of corresponding LOSs.

*4.4 GIX, NeGIX and TEGIX indices*

To cross-validate the findings of the analysis of the S4, ROTI and $\nabla$TEC data presented above, we used the GIXy component (GNSS based gradient ionosphere index) and NeGIX and TEGIX (Swarm data-based indices) data. Figure 22 shows half-hourly maps of 95 percentile of the GIXy (gradient in geographic North-South direction) component, from 20:30 UTC of June 22 until 02:00 UTC of June 23. These maps clearly show clear signatures of the dynamic evolution of TEC gradients in agreement with other results aforementioned (see Figs. 16 and 22). Regions indicated by purple circles on June 23 during 00:00 – 02:00 UTC show presence of positive (i.e., higher TEC for the Northern IPPs) and negative (i.e., higher TEC for the Southern IPPs) GIX indicating signatures of high dynamical regions such as IPBs. The provision of sign in GIX provides a unique opportunity of identifying such regions. The results agree with the Swarm PBI detections (vertical lines of grey dots) in Fig. 18 for June 22 23:00 and June 23 01:00 UTC. GIXx (gradient in West-East direction) component does not show significant changes in gradient maps and therefore is not included. It must be mentioned that the observed GIX related TEC gradients have mid-to-large scale character in agreement with [8]. Since S4 and ROTI data indicate small to mid-scale irregularities, it can be concluded that the ionosphere over the South part of the Iberian Peninsula was perturbed at all spatial scales on June 22, 2015. This conclusion is confirmed by gradient estimates from Swarm data on June 22, 2015.

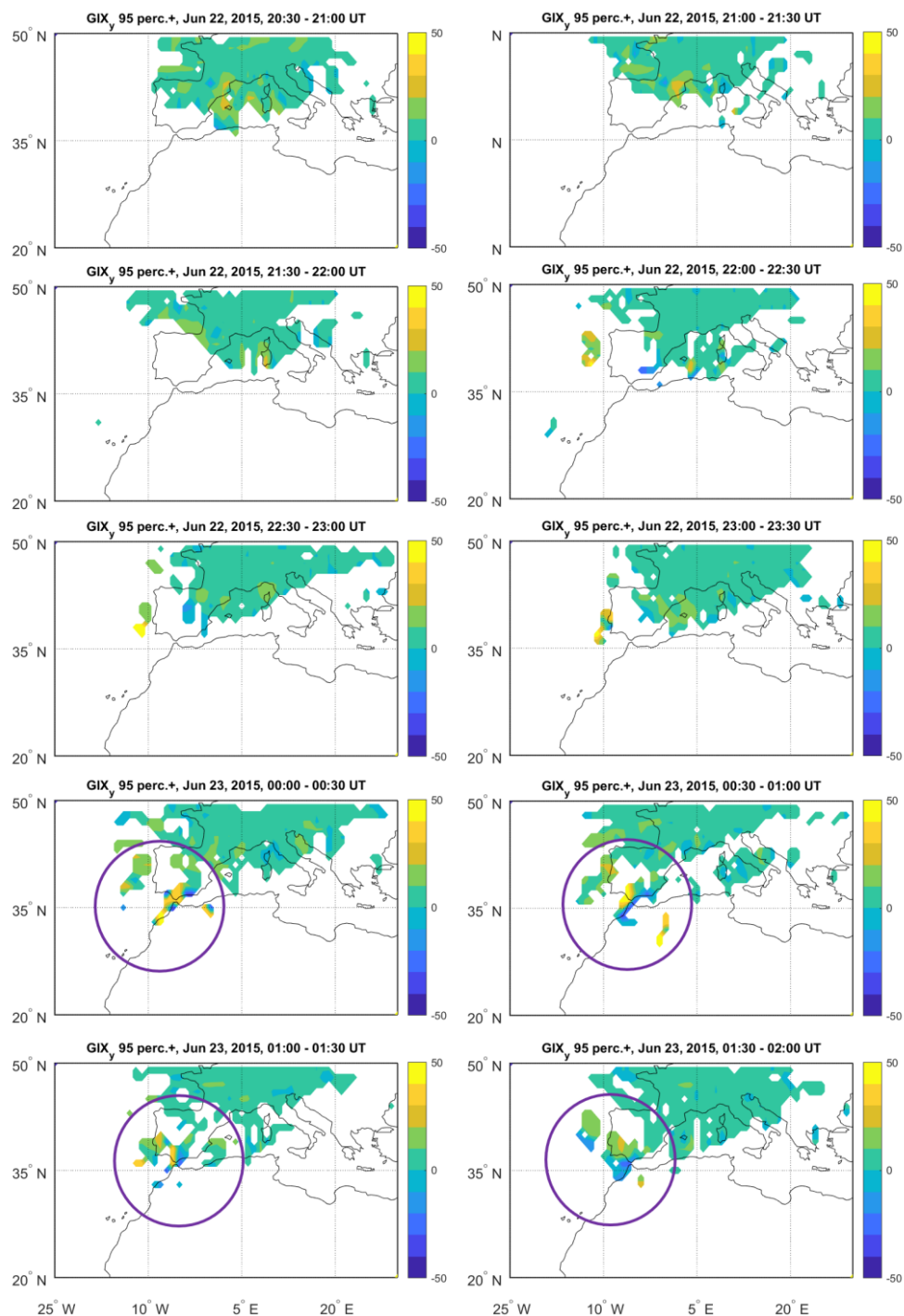

**Figure 22.** Maps of 95=percentile of the GIXy component (i.e., -ve means higher gradient in South-North direction) in mTECU/km averaged over every 30 minutes interval from 20:30 UTC of June 22 until 02:00 UTC of June 23. The dipole length range (50–1,000 km) is rather large to obtain sufficient data coverage for mapping. Regions indicated by circles show existence of +ve and -ve gradients indicating regions of high gradients.

The NeGIX and TEGIX indices are calculated using the Swarm observations. Figures 23 (top) and 24 display the resulting electron density gradients for the orbits of Swarm A and C that navigate Westwards from the evening of June 22 to early hours of June 23, 2015.

As previously, the data of only two passes of Swarm on June 22 were used for the analysis: at 21:32 UTC (along the 18° E meridian) and at 23:06 UTC (along the 5.5° W meridian). As one can see from Fig. 23, these orbital passes are the ones with the strongest Ne and TEC gradients during the studied time interval. The location of these identified ionospheric perturbations in the region of the Strait of Gibraltar and Northern Africa correlates very well with other observational data presented above. As seen from Figure 24, the most extreme 95-percentile values of the electron density gradients for these two orbits are seen between 20° N and 40° N and surpassed 5000 e/cm$^3$·km for the zonal and meridional components of NeGIX. This value is 10x greater than for the other 3 orbits of Swarm A and C.

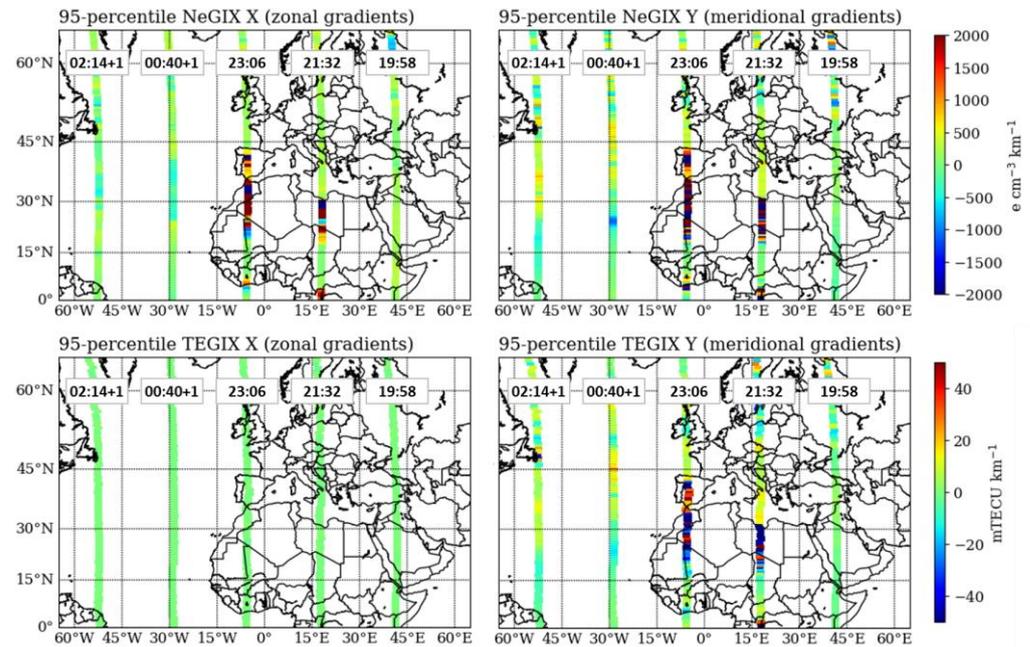

**Figure 23.** Resulting maps for the 95-percentile values of the X and Y components of NeGIX (top), and TEGIX (bottom). The indices are determined using data from Swarm A and C, that orbit from the right pass in the evening of June 22 towards the left in the early hours of June 23. The paths at 21:32 and 23:06 UTC show the strongest gradients for this event and are reviewed in more detail below.

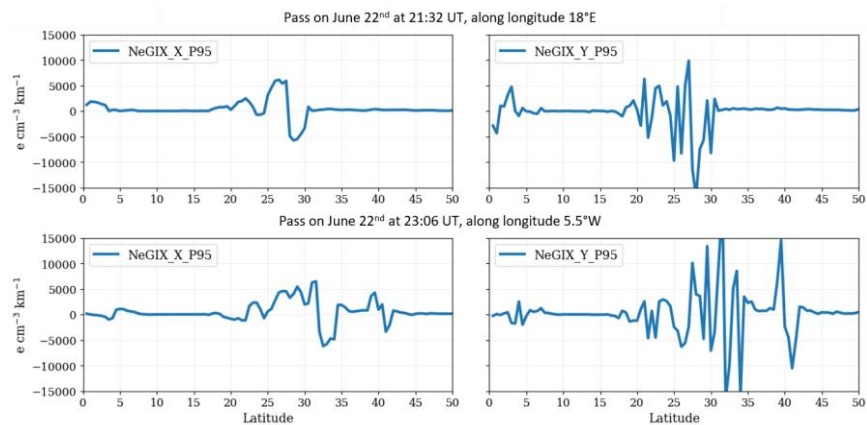

**Figure 24.** 95-percentile values for the X (left) and Y (right) components of NeGIX as a function of latitude. As seen from the maps above, the two orbits of Swarm A and C with the strongest gradient values on June 22 are traced from top to bottom.

Also, the meridional component of TEGIX on the right panels of Figure 25 exposes strong 95-percentile values of TEC gradients over the same latitudinal range, reaching absolute values of more than 100 mTECU/km. For comparison, with red dashed lines the X and Y components of the spatial gradients extracted from ∇TEC maps (GNSS data) are overplotted as a function of latitude. ∇TEC results are limited to latitudes above 30 N. Nevertheless, this comparative analysis with the values of TEGIX for the orbit at 21:32 UTC shows consistent results in magnitude. For the pass at 23:06 UTC, although strong TEC variability in the region is detected by the two methods, a strong difference in amplitude of gradients is observed. This can be explained by the natural definition of these approaches - whereas the estimation of ∇TEC gradients is a simplified approach to compare TEC values at adjacent grid points of TEC maps, the richer sample of Swarm A and C TEC measurements and their combination to form gradient vectors permit to deduce a 95-percentile. In this context, Swarm permits to overcome limitations of data availability and resolution as generally experienced with ∇TEC maps and ground-based GNSS techniques, especially in the studied region.

Still, the contrast between ground-based and space-born instruments serves to confirm the location of ionospheric inhomogeneities and the dynamics of the studied parameters.

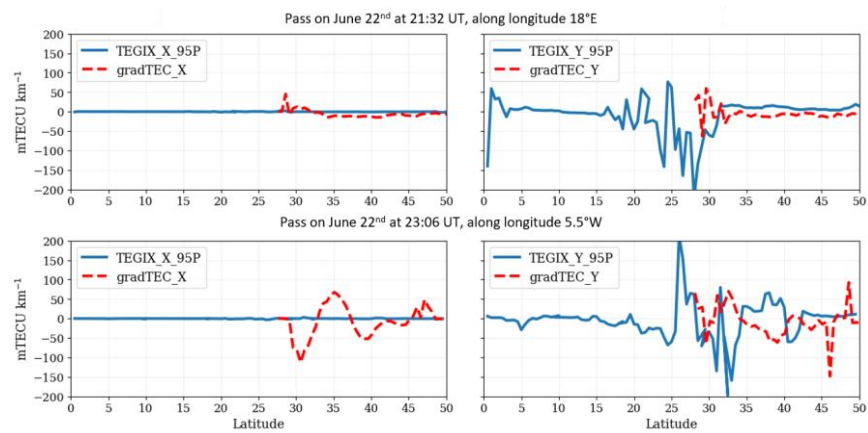

**Figure 25.** Comparison between the 95-percentile values for the X and Y components of TEGIX (blue lines) and the spatial gradients for similar meridians extracted from ∇TEC maps (red dashed lines) as a function of latitude. The two orbits of Swarm A and C with the strongest gradient values are displayed at 21:32 UTC (top) and 23:06 UTC (bottom).

## 5. Discussion and Conclusions

Monitoring ionospheric irregularities over the Western Mediterranean is of paramount importance due to the potential for small-scale irregularities to spill over into midlatitudes, especially during disturbed Geospace conditions. In the case event of the June 2015 storm, we contributed to the understanding of the conditions that lead to the penetration into these latitudes and pinpoint how their monitoring and characterization are crucial, given the growing reliance on GNSS services and technologies, which are highly susceptible to disruptions caused by those kinds of threatening ionospheric irregularities. In fact, the continuous monitoring and analysis of ionospheric irregularities in this region (which is located in the South and South-West part of the area covered by EGNOS, a satellite-based augmentation system developed by the European Space Agency and EUROCONTROL to enhance the reliability and accuracy of the positioning data) are essential to mitigate these risks and ensure the integrity of critical GNSS-dependent technologies.

The dynamics of the ionosphere in the low-middle latitudes of the Euro-African meridional sector during the geomagnetic storm of June 22-23, 2015, was different from what is normally expected for this region. Large variations of scintillation indices (S4 and ROTI) as well as large spatial TEC gradients were observed there between 19 h UTC of June 22 and 03 h UTC of June 23. These ionospheric disturbances were caused by a rare event of a spill-over of EPBs from low latitudes triggered by the storm dynamics.

During the studied event, EPBs were seen as elongated (NW-SE) areas of lower electron density (lower TEC) travelling North-Westward through the studied region. High values of S4 and ROTI obtained from individual GNSS receivers were located, as a rule, inside those areas.

The spatial and temporal variations of different parameters such as scintillation indices, TEC gradients and TEC and electron density indices, obtained from the ground-based and space-born instruments, are consistent, and clearly show that spill-over EPBs are responsible for specific ionospheric conditions that were observed during this storm in this meridional sector.

The unusual character of the ionospheric response to the storm of June 2015 was already noted in [8] when compared to the geomagnetic storm of March 2015. During the storm of March 2015, the strength of the ionospheric disturbance (measured by the GIX and SIDX indices, and ROTI) gradually decreased from high (60° - 75° N) to low-middle (30° - 45° N) latitudes. This observation indicates a growing dissipation of travelling ionospheric disturbances generated by the solar energy input at high latitudes during the storm and subsequently moving equatorward. On contrary, as is shown in [8], the highest values of GIX were observed both at high and low-middle latitudes, and even the ROTI values obtained for the low-middle latitudes were higher than for the middle (45° - 60° N) latitudes. The results of a study presented in this paper allows assuming that the difference in the latitudinal response of the ionosphere to these two storms of 2015 results from the EPB spill-over event. The rapid changes of spatial gradients and positioning errors shown in [8] can also be explained by the effect of passing spilled-over EPBs. It is interesting to note that GIX indicates medium to large scale (here 50-500km) perturbations of TEC. Thus, a broad spectrum of spatial and temporal scales has been observed in the Western Mediterranean area on June 22, 2015. The related physical processes originate probably in lower latitudes.

In conclusion, this work contributes to a deeper understanding of the longitudinal variations in the North-African sector, particularly in the Western region where EPB occurrences are more intense. These variations, whose origin is not completely understood, play a key role in determining the severity and frequency of EPB spill-over, and their study is willing to contribute to future more accurate forecasting and mitigation strategies. Lastly, this work can be of support towards the planning of enhanced monitoring infrastructure and research efforts in the Western Mediterranean to address the challenges posed by ionospheric irregularities and ensure the continued reliability of GNSS services in the region.

**Supplementary Materials:** The following supporting information can be downloaded at: www.mdpi.com/xxx/s1,

*Figures*: Figs_S01_S02.pdf: S4 multipath analysis for Lampedusa and Tenerife, respectively (same as Fig. 2); *Animations:* Anim_S01.gif: timelapse of S4 IPP maps between June 22 19:30 and June 23 02:30; Anim_S02.gif: timelapse of ROTI IPP maps between June 22 19:30 and June 23 02:30; Anim_S03.gif: timelapse of ROTI maps between June 22 19:30 and June 23 02:30; Anim_S04.gif: timelapse of ROTI and S4 IPP maps between June 22 19:30 and June 23 02:30; Anim_S05.gif, Anim_S06.gif and Anim_S07.gif: timelapse of absolute, x and y, respectively, TEC gradient maps and ROTI between June 22 19:30 and June 23 02:30; Anim_S08.gif, Anim_S09.gif and Anim_S10.gif: timelapse of absolute, x and y, respectively, TEC gradient maps and S4 between June 22 19:30 and June 23 02:30;.

**Author Contributions:** This work has the following contributions:

conceptualization—AM;

methodology—AM, LS, TB;

software—AM, RI, EP, DE;

validation—AM, LS, TB;

formal analysis—AM;

investigation—AM, LS, TB, RI, EP, MH, AC, DE;

resources—AM, TB, LS;

data curation—AM, RI, EP;

writing—original draft preparation—AM;

writing—review and editing—AM, LS, TB, RI, EP, MH, AC, NJ;

visualization—AM, AC, MH;

supervision—AM;

project administration—AM;

funding acquisition—AM, LS

All authors have read and agreed to the published version of the manuscript.

**Funding:** IA is supported by Fundação para a Ciência e a Tecnologia (FCT, Portugal) through the research grants UIDB/04434/2020 and UIDP/04434/2020. This study is a contribution to the "PROSE: P3-SWE-XXXVII - SWE PRODUCTS FOR SOUTHERN EUROPE - PHASE 1" project funded by ESA. The work was supported through the "ALERT: Assessment of the ionospheric scintillation over Portugal" project as a part of the 3rd PITHIA-NRF TNA. The PITHIA-NRF project has received funding from European Union's Horizon 2020 research and innovation program under grant agreement no. 101007599. AM acknowledges the research infrastructure and the access provider INGV of the PITHIA-NRF project (https://www.pithia-nrf.eu/). RI's current research fellowship is funded by the Swarm Space Weather Variability, Irregularities, and Predictive Capabilities for the Dynamic Ionosphere (Swarm-VIP-Dynamic) project, which has been funded by the European Space Agency, contract 4000143413/23/I-EB within the "Esa Solid Magnetic Science Cluster - Research Opportunities: 4dionosphere - Expro+".

**Data Availability Statement:** The SCINDA data are available: for [21] at Barlyaeva, T.; Barata, T.; Morozova, A. Data from: Datasets of ionospheric parameters provided by SCINDA GNSS receiver from Lisbon airport area. Mendeley Data 2020, V1. http:/dx.doi.org/10.22632/kkytn5d8yc.1; for [22] at Morozova, A.; Barlyaeva, T.; Barata, T. Data from: Datasets of ionospheric parameters (TEC, SI, positioning errors) from Lisbon airport area for 2014–2019. Mendeley Data 2022, V2. http://dx.doi.org/10.22632/3z6mjk39jv.2.

The RENEP RINEX 2.11 files are available through https://renep.dgterritorio.gov.pt/ (accessed on November 18, 2024).

Rinex data from the mas1, lpal, rabt and meli receiver are accessible through the FTP repository of the BKG's GNSS Data Center (https://igs.bkg.bund.de/), part of the International GNSS Service (IGS).

Data for the Novatel ISMR receiver operating in Lampedusa (code "lam0s") are accessible through the eSWua repository (https://doi.org/10.13127/eswua/gnss).

The GNSS data for the Tenerife station are by courtesy of the Ionosphere Monitoring and Prediction Center at DLR, Germany (https://impc.dlr.de/)

TEC and ROTI maps are from the DRAWING-TEC project by ISEE (https://aer-nc-web.nict.go.jp/GPS/DRAWING-TEC/) and available at https://stdb2.isee.nagoya-u.ac.jp/GPS/GPS-TEC/ and https://stdb2.isee.nagoya-u.ac.jp/GPS/GPS-TEC/GLOBAL/RMAP/index.html (last visited on *** 202*). Global GNSS-TEC data processing has been supported by JSPS KAKENHI Grant Number 16H06286. GNSS RINEX files for the GNSS-TEC processing are provided from many organizations listed by the webpage.

(http://stdb2.isee.nagoya-u.ac.jp/GPS/GPS-TEC/gnss_provider_list.html).

Ionospheric plasma bubble index is from the SW_IBIxTMS_2F Swarm product (please visit https://swarmhandbook.earth.esa.int/catalogue/sw_ibixtms_2f for description) and available at ftp://swarm-diss.eo.esa.int/Level2daily/Latest_baselines/IBI/TMS (last visited on \*\*\* 202\*).

**Acknowledgments:** Authors are grateful to Martin Kriegel (DLR) for providing Tenerife data. Authors would like to acknowledge the Direção Geral do Território (DGT) and Helena Ribeiro personally for making ReNEP data available (RENEP). Authors thank Giorgio di Sarra, Scientific director of the activities at the ENEA Climate Observation Station, and Damiano Sferlazzo for their support with the ISMR station in Lampedusa.

**Conflicts of Interest:** The authors declare no conflicts of interest. The funders had no role in the design of the study; in the collection, analyses, or interpretation of data; in the writing of the manuscript; or in the decision to publish the results.